\documentclass[12pt,onecolumn,draftclsnofoot]{IEEEtran}

\usepackage{times}
\usepackage{amsmath, amsthm}
\usepackage{amsfonts}
\usepackage{mathrsfs}
\usepackage{latexsym}
\usepackage{amssymb}
\usepackage{enumerate}
\usepackage{upref}
\usepackage{graphicx}
\usepackage{color}
\usepackage{scrextend}
\usepackage{algorithm}
\usepackage{algpseudocode}
\usepackage{verbatim}
\usepackage{multicol}
\usepackage{colortbl}
\usepackage{multirow}
\usepackage{makecell}
\usepackage{mathrsfs}
\definecolor{mygray}{gray}{.9}
\usepackage{cite}
\usepackage{xcolor}

\usepackage{textcomp}
\usepackage[paper=letterpaper,top=1.1in,bottom=0.8in,left=0.8in,right=0.8in]{geometry}

\theoremstyle{definition}

\newtheorem{prop}{Proposition}
\newtheorem{thm}{Theorem}
\newtheorem{cor}[prop]{Corollary}
\newtheorem{lem}{Lemma}

\newtheorem{cnstr}[prop]{Construction}

\newtheorem{exa}{Example}

\newtheorem{rem}{Remark}

\begin{document}
	

	\title{Flexible Distributed Matrix Multiplication}
	
	\author{ 
		\IEEEauthorblockN{Weiqi Li, Zhen Chen, Zhiying Wang, Syed A. Jafar, Hamid Jafarkhani}\\
		\IEEEauthorblockA{Center for Pervasive Communications and Computing (CPCC)  \\ University of California, Irvine, USA
			\\ \{weiqil4, zhenc4, zhiying, syed, hamidj\}@uci.edu
		} 
		\thanks{This paper was presented in part at ISIT 2021 \cite{Li_flexible}.}
	}

	\maketitle

\begin{abstract}
	The distributed matrix multiplication problem with an unknown number of stragglers is considered, where the goal is to efficiently and flexibly obtain the product of two massive matrices by distributing the computation across $N$ servers. There are up to $N - R$ stragglers but the exact number is not known a priori. Motivated by reducing the computation load of each server, a flexible solution is proposed to fully utilize the computation capability of available servers. The computing task for each server is separated into several subtasks, constructed based on Entangled Polynomial codes by Yu et al. The final results can be obtained from either a larger number of servers with a smaller amount of computation completed per server or a smaller number of servers with a larger amount of computation completed per server. The required finite field size of the proposed solution is less than $2N$.  Moreover, the optimal design parameters such as the partitioning of the input matrices is discussed. Our constructions can also be generalized to other settings such as batch distributed matrix multiplication and secure distributed matrix multiplication.
\end{abstract}


\section{Introduction}
Distributed matrix multiplication has received wide interest because of the huge amount of data computation required by many popular applications like federated learning, cloud computing, and the Internet of things. In particular, the multiplication of two massive input matrices $A\in \mathbb{F}^{\lambda \times \kappa}$ and $B\in \mathbb{F}^{\kappa \times \mu}$, where $\mathbb{F}$ is some finite field is considered. Each matrix is encoded into $N$ \emph{shares} and distributed to $N$ servers. Each server performs computation on its own shares and sends the \emph{results} to the central computational node, e.g., the cloud. After collecting enough results, the desired product $AB$ can be calculated. However, stragglers (servers that fail to respond or respond after the the reconstruction is executed) are inevitable in distributed systems, due to various reasons \cite{Dean_tail, Anan_straggler} including network latency, resource contention, workload imbalance, failures of hardware or software, etc. To reduce the overall system latency caused by stragglers, distributed matrix computing schemes with straggler tolerance are provided in \cite{Lee_Lam_Pedarsani, Yu_Maddah-Ali_Avestimehr_Polynomial,Dutta_Fahim_Haddadpour, GPolyDot, Yu_Maddah-Ali_Avestimehr, Yu_Lagrange, Reisizadeh_Prakash_Pedarsani, Lee_Suh_Ramchandran, Dutta_Cadambe_Short, Dutta_Cadambe_Codedconv, Yu_Maddah-Ali_CodedDFT, Jahani-Nezhad_Maddah-Ali, Baharav_Lee_Ocal, Suh_Lee_Msparse, Wang_Liu_CLT, Mallick_Chaudhari_Joshi, Wang_Liu_Sparse, Severinson_iAmat_Rosnes, Haddadpour_Cadambe_Finite,Sheth_Dutta_Chaudhari, Jeong_Ye_Grover,Kim_Sohn_Moon_Group,Park_Lee_Sohn,Li_Maddah-Ali_Fog, Chang_Tandon, Kakar_Ebadifar_Sezgin_CSA, Oliveira_Rouayheb_Karpuk, Kim_Lee, Aliasgari_Simeone_Kliewer, Jia_Jafar_CDBC, Chen_Jia_Wang_Jafar_GCSANA} 
 with a predetermined \emph{recovery threshold} $R$ such that the final product can be obtained using computation results from any $R$ out of $N$ servers.
Among the state-of-the-art schemes, some are based on matrix partitioning such as Polynomial codes \cite{Yu_Maddah-Ali_Avestimehr_Polynomial}, MatDot codes and PolyDot codes \cite{Dutta_Fahim_Haddadpour}, Generalized PolyDot codes \cite{GPolyDot} and Entangled Polynomial (EP) codes \cite{Yu_Maddah-Ali_Avestimehr}, and others are based on batch processing such as Lagrange Coded Computing \cite{Yu_Lagrange}, Cross Subspace Alignment (CSA) codes and Generalized Cross Subspace Alignment (GCSA) codes \cite{Jia_Jafar_CDBC}. 

The above literature assumes there are a fixed number $N-R$ of stragglers. However, the number of stragglers is unpredictable in practical systems. When the number of stragglers is smaller than $N-R$, each non-straggler server still needs to do the same amount of computation as if there are $N-R$ stragglers and the central node still only uses the results from $R$ servers. A significant amount of computation power is wasted. To handle this situation, a setting in which the number of stragglers is not known a priori has been considered in \cite{Nuwan_Stark_ISIT, Amiri_cdc, Bitar_Parag_Rouayheb_staircasecodes, Bitar_Xing_adaptive, Ramamoorthy_universal, Das_c3les, Bitar_Xhemrishi_adaptive, Kiani_Nuwan_Stark_ISIT, Hasircioglu_bivariate, Hasircioglu_Hermitian, Fan_partial, Das_sparse, Shahrzad_Hierarchical}
and schemes that can cope with such a  setting have been designed. The underlying idea is to assign a sequence of small tasks to each server instead of assigning a single large task. Therefore, besides the scenario that the fastest $R$ servers finish all their tasks, there are other scenarios that make the computation complete. References \cite{Nuwan_Stark_ISIT, Amiri_cdc} focus on the task scheduling for general distributed computing. The matrix-vector multiplication setting is considered in \cite{Bitar_Parag_Rouayheb_staircasecodes, Bitar_Xing_adaptive, Ramamoorthy_universal, Das_c3les}. In these works, only the input matrix is partitioned. References \cite{Bitar_Xhemrishi_adaptive, Kiani_Nuwan_Stark_ISIT, Hasircioglu_bivariate, Hasircioglu_Hermitian, Fan_partial, Das_sparse} consider matrix-matrix multiplication, but they can only handle a special partitioning, i.e., $A$ and $B$ are row-wisely and column-wisely split, respectively, or only $A$ is row-wisely split. In \cite{Shahrzad_Hierarchical}, the authors propose $3$ hierarchical schemes for matrix multiplication to leverage partial stragglers. The main idea is that the task is first divided into several small subtasks, i.e., the multiplication of several pairs of small matrices, and each subtask is coded separately with existing schemes. 

Arbitrary partitioning of input matrices is important in massive matrix multiplication since it enables different utilization of system resources, e.g., the required amount of storage at each server and the amount of communication from servers to the central node. When the number of stragglers is fixed, many codes such as PolyDot codes \cite{Dutta_Fahim_Haddadpour}, EP codes \cite{Yu_Maddah-Ali_Avestimehr} and GCSA codes \cite{Jia_Jafar_CDBC} provide elegant solutions for arbitrary partitioning by encoding the input matrix blocks into a carefully designed polynomial. In particular, EP codes 
effectively align the servers’ computation with the terms that the central node needs and achieve the optimal recovery threshold among all linear coding strategies in some cases. 

A naive solution to achieve flexibility for distributed matrix multiplication with arbitrary partitioning is simply applying a fixed EP code with a recovery threshold of $PR$, where each server gets $P$ pairs of shares instead of one pair of shares. The central node can calculate the final results with any $PR$ out of the $PN$ computing results. Thus, each server only needs to compute $RP/N$ results when there is no straggler, and in general, the number of results computed in each server can be adjusted based on the number of stragglers. However, by doing so, the computation needs to be done in a field with a minimum size of $PN$ and operations over a larger field result in a much bigger delay \cite{gashkov2013complexity}. 

In this paper, we present a flexible coding scheme for distributed matrix multiplication that allows a flexible number of stragglers and arbitrary matrix partitioning while only requiring a much smaller field size. The main idea is that non-straggler servers can finish more tasks to compensate for the effect of the stragglers without knowing the stragglers a priori. Specifically, the computation is encoded into several tasks for each server, and each server keeps calculating and sending results to the central node until enough results are obtained. Enough results can be either a larger number of servers with a smaller amount of completed computation by each server or a smaller number of servers with a larger amount of completed computation by each server. Therefore, the number of available servers is flexible and the number of required tasks is adjusted to the number of available servers. Our scheme is different from those that leverage partial stragglers \cite{Kiani_Nuwan_Stark_ISIT, Fan_partial, Ramamoorthy_universal, Das_sparse, Das_c3les, Shahrzad_Hierarchical}. In our construction, the computation load (the number of multiplication operations) of each non-straggler server is the same and the computation by stragglers is neglected, while in schemes with partial stragglers, the computation load varies in different servers including stragglers. 


The main contributions of the paper are as follows. We present a coding framework of flexible distributed matrix multiplication schemes, and one-round and multi-round communication models.
A construction with multiple layers of computation tasks is proposed, which only requires a field size of less than $2N$ and the computation load of each server is reduced significantly when there are fewer stragglers than $N-R$. We also demonstrate the optimization of the parameters to obtain the lowest computation load. We show that the two-layer construction outperforms the fixed scheme under the one-round model as long as the server storage is above a threshold, and the maximum number of layers is preferred under the multi-round model.

The rest of the paper is organized as follows. Section \ref{problem} presents the problem statement. In Section \ref{construction}, we present our construction and its performance. The choice of parameters to optimize the computation load given the storage capacity is discussed in Section \ref{optimization}. Section \ref{conclusion} concludes the paper.

\emph{Notation:} We use calligraphic characters to denote sets. 
For positive integer $N$, $[N]$ stands for the set $\{1,2,\dots,N\}$. 
For a matrix $M$, $|M|$ denotes its number of entries. 
For a set of matrices $\mathcal{M}$, $|\mathcal{M}|$ represents the sum of the number of entries in all its matrices.
When $M$ is partitioned into sub-block matrices, $M_{(u,v)}$ denotes the block in the $u$-th row and the $v$-th column.

\section{Problem Statement} \label{problem}
\begin{figure}
    \centering
    \includegraphics[width=0.5\textwidth]{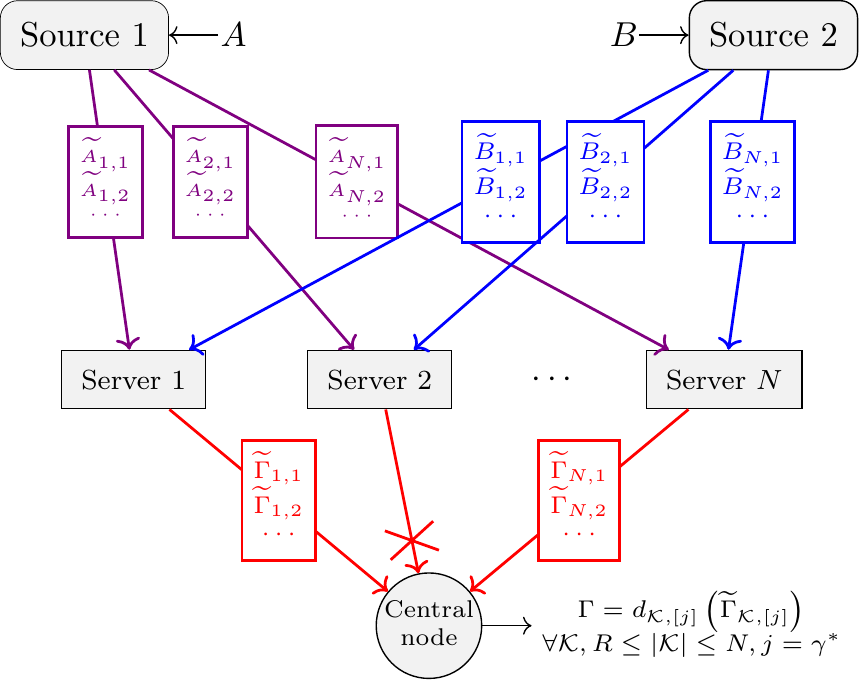}
    \caption{The flexible distributed matrix multiplication problem.}
    \label{fig:problem}
\end{figure}

We consider a problem of matrix multiplication (see Fig. \ref{fig:problem}) with two input matrices $A\in \mathbb{F}^{\lambda \times \kappa}$ and $B\in \mathbb{F}^{\kappa \times \mu}$, for some integers $\lambda, \kappa, \mu$ and a field $\mathbb{F}$. We are interested in computing the product $\Gamma = AB$ in a distributed computing environment with $2$ sources, a central node, and $N$ servers. Sources $1$ and $2$ hold matrices $A$ and $B$, respectively. It is assumed that there are up to $N-R$ stragglers among the servers. In non-flexible distributed matrix multiplication, $R$ is called the \emph{recovery threshold}. 
The shares (coded matrix sets) $\widetilde{\mathcal{A}}_i$ and $\widetilde{\mathcal{B}}_i$ are generated by sources for Server $i,i\in[N]$. Each share consists of some coded matrices, denoted by $\left\{\widetilde{{A}}_{i,1}, \cdots, \widetilde{{A}}_{i,\widetilde{\gamma}} \right\}$, or $\left\{\widetilde{{B}}_{i,1}, \cdots, \widetilde{{B}}_{i,\widetilde{\gamma}} \right\}$, where $\widetilde{\gamma}$ is a function of $N$ and $R$. For $i\in[N]$, the shares
and the encoding functions are
\begin{align}
\widetilde{\mathcal{A}}_i = \left\{\widetilde{{A}}_{i,j} \mid j \in [\widetilde{\gamma}] \right\} = u_i(A),  \\
\widetilde{\mathcal{B}}_i = \left\{\widetilde{{B}}_{i,j} \mid j \in [\widetilde{\gamma}] \right\} = v_i(B).
\end{align}
Then, $\widetilde{\mathcal{A}}_{i}$ and $\widetilde{\mathcal{B}}_{i}$ are sent to Server $i$ from the sources. 
Server $i$ sequentially computes $\widetilde{\gamma}$ \emph{tasks} in order:
\begin{align}
\widetilde{{\Gamma}}_{i, j} = \widetilde{{A}}_{i,j} \cdot \widetilde{{B}}_{i,j}, \text{ } j \in [\widetilde{\gamma}],
\end{align}
and sends $\widetilde{{\Gamma}}_{i, j}$ to the central node once its computation is finished. Due to the sequential processing nature of the servers, 
the central node receives $\widetilde{\Gamma}_{i, j_1}$ before $\widetilde{{\Gamma}}_{i, j_2}$ for $\forall i \in [N], j_1 < j_2$. Denote $\widetilde{\Gamma}_{i, [j]} = \left \{ \widetilde{{\Gamma}}_{i, t} \mid t\in[j]  \right\}$ and $\widetilde{{\Gamma}}_{\mathcal{K}, [j]} = \left\{ \widetilde{{\Gamma}}_{i, [j]} \mid i \in \mathcal{K}\right\}, \forall \mathcal{K} \subset [N] $. 

We require that the central node be able to decode the desired product $\Gamma$ from arbitrary $\hat{R} \ge R$ servers, where each server calculates $\gamma^\ast$ (a function of $\hat{R}$) tasks. Equivalently, the decoding function $d_{\mathcal{K}, [j]}$ of the central node for recovering $\Gamma$ satisfies
\begin{align}
\Gamma = d_{\mathcal{K}, [j]}\left( \widetilde{{\Gamma}}_{\mathcal{K}, [j]} \right),  \forall \mathcal{K},  R \le |\mathcal{K}|=\hat{R}  \le N, j = \gamma^\ast.
\end{align}

The function set $\{u_i, v_i, d_{\mathcal{K}, [j]} \mid 1 \le i \le N, R \le |\mathcal{K}| =\hat{R} \le N, j = \gamma^\ast\}$ is called the \emph{flexible constructions for distributed matrix multiplication}.

In other words, the sources send coded matrices to each server. Each server keeps calculating and sending results to the central node until it obtains enough results -- when the quickest $\hat{R}$ servers complete the first $\gamma^\ast$ tasks. The remaining servers are viewed as stragglers and the computation results from stragglers are ignored.  

In this work, we consider two communication models: the one-round communication model and the multi-round communication model. For the \emph{one-round communication model}, the sources send \emph{all} $\widetilde{\gamma}$ coded matrices to the server at one time. After that there are no communications between sources and servers. For the \emph{multi-round communication model}, first, the sources send one pair of coded matrices to the servers. Once a server finishes its tasks, it will ask the sources to send another pair of coded matrices. It is not necessary for the sources to know which servers are the stragglers. This procedure lasts until the central node obtains enough results. Note that there are no communications among servers in either models.

The \emph{computation load} $L$ is defined as the number of multiplication operations per server. Moreover, each server has a \emph{storage capacity} $C$ \footnote{The maximum storage size $C$ is usually smaller than $|A|+|B|$, otherwise the sources can send $A$ and $B$ to the servers.}. At any time, any server cannot store more than $C$. Specifically, for the one-round communication model, $\max_{i \in [N]} \left(\left\lvert\widetilde{\mathcal{A}}_{i}\right\rvert + \left\lvert\widetilde{\mathcal{B}}_{i}\right\rvert \right) \le C$. For the multi-round communication model, $\max_{i \in [N], j \in [\widetilde{\gamma}]} \left(\left\lvert\widetilde{A}_{i,j}\right\rvert + \left\lvert\widetilde{B}_{i,j}\right\rvert \right) \le C$.
This is because once a server finishes a task and sends the result to the central node, it can refresh the storage and delete the coded matrices related to this task. In general, the storage constraint is stricter in the one-round communication model. 
We want to find flexible constructions with the \emph{storage capacity} $C$ and the \emph{computation load} $L$ at each server as small as possible.

\section{Construction} \label{construction}
In this section, we present our flexible constructions. The scheme is based on EP code \cite{Yu_Maddah-Ali_Avestimehr} and the computation tasks are divided into several layers to provide flexibility. We start with a motivating example. The general construction and its storage and computation load are presented afterwards.

\begin{exa}\label{example 1}
Consider the matrix multiplication of $A$ and $B$,  for $A \in \mathbb{F}^{\lambda \times \kappa}, B \in \mathbb{F}^{\kappa \times \mu}$, using $N=5$ servers with at most $N-R=2$ stragglers. Suppose $A$ is column-wisely partitioned as $A = [A_1, A_2]$, each submatrix is of size $\lambda\times\frac{\kappa}{2}$, and $B$ is row-wisely partitioned as $B =\left[\begin{array}{c} B_1  \\ B_2 \\ \end{array}\right]$, each submatrix is of size $\frac{\kappa}{2}\times\mu$. The central node requires $AB = A_1B_1 + A_2B_2$. Applying the EP code, Server $i,i\in[5]$ receives coded matrices
$A_1 + \alpha_i A_2$ and $\alpha_i B_1+B_2$, for $\alpha_i \in \mathbb{F}$,
and calculates
\begin{align}\label{EP example}
	&(A_1 + \alpha_i A_2)\cdot(\alpha_i B_1+B_2) \\ \nonumber =& A_1B_2 + \alpha_i (A_1B_1 + A_2B_2) + \alpha_i^2 A_2B_1,
\end{align}
which is a degree $2$ polynomial with respect to $\alpha_i$. Thus, $A_1B_1 + A_2B_2$ can be calculated by $3$ distinct evaluations from $\{\alpha_i \mid i\in [5]\}$ using Lagrange interpolation. The total computation load of directly multiplying $A$ and $B$ is $L=\lambda\kappa\mu$, while using the EP code the computation load of each server is $L/2$. However, when there is no straggler, the computation of $2$ servers are wasted.

Alternatively, we can use a flexible scheme to calculate $AB$, such that any $\hat{R}$ available servers can complete the computation, $3=R \le \hat{R} \le N=5$. First, we partition the matrices and get $A = [A_1, A_2, A_3]$, each submatrix is of size $\lambda\times\frac{\kappa}{3}$, and $B = [B_1^T,   B_2^T,  B_3^T]^T $, each submatrix is of size $\frac{\kappa}{3}\times\mu$. The central node requires $AB=A_1B_1 + A_2B_2 + A_3B_3$. 
Let $\{\alpha_i | i \in [7]\}$ be distinct elements in $\mathbb{F}$.
The calculation will be divided into $2$ layers.

Layer 1: Server $i,i\in[5],$ calculates $\gamma_1 = 1$ task
\begin{align}
	\Gamma_{i,1} =&(A_1 + \alpha_i A_2 + \alpha_i^2 A_3)\cdot(\alpha_i^2 B_1 +\alpha_i B_2+B_3) \nonumber\\ 
	=& A_1B_3 + \alpha_i (A_2B_3 + A_1B_2) + \alpha_i^2( A_1B_1 + A_2B_2 + A_3B_3) \nonumber\\ 
	& + \alpha_i^3 (A_2B_1 + A_3B_2) + \alpha_i^4 A_3B_1.
\end{align}
It is a degree $4$ polynomial with respect to $\alpha_i$ and the final product can be obtained from all $5$ servers. If there is no straggler, we stop here. In this layer, matrices $A$ and $B$ are divided into smaller pieces compared to the fixed EP code and the computation load of each server is $L/3$. If there are stragglers, the servers continue the calculation in Layer 2.

Layer 2: We set $A_{\alpha_i} = A_1 + \alpha_i A_2 + \alpha_i^2 A_3, B_{\alpha_i} = \alpha_i^2 B_1 + \alpha_i B_2 + B_3, i\in\{6,7\}$ and partition them into $2$ parts,
\begin{align}
	A_{\alpha_i} = [A_{\alpha_i,1}, A_{\alpha_i,2}], B_{\alpha_i} = \left[\begin{array}{c} B_{\alpha_i,1}  \\ B_{\alpha_i,2} \\ \end{array}\right].
\end{align}
Server $i$ has $\gamma_2=2$ computation tasks: 
\begin{align}
	\Gamma_{i,2} &= (A_{\alpha_6,1}+ \alpha_i A_{\alpha_6,2}) \cdot (\alpha_i B_{\alpha_6,1}+  B_{\alpha_6,2}), \\
	\Gamma_{i,3} &= (A_{\alpha_7,1}+ \alpha_i A_{\alpha_7,2}) \cdot (\alpha_i B_{\alpha_7,1}+  B_{\alpha_7,2}).
\end{align}
The detailed calculation of each server is shown in Table \ref{table: example 1}. 

Since Layer 2 has a similar structure as \eqref{EP example}, from any $3$ of the servers, we can get $A_{\alpha_6} \cdot B_{\alpha_6}$ and/or $A_{\alpha_7} \cdot B_{\alpha_7}$. If there is one straggler, the central node obtains $A_{\alpha_6} \cdot B_{\alpha_6}$ from Layer $2$, which causes the additional computation load  of $L/6$ in a server. If there are $2$ stragglers, the central node obtains both $A_{\alpha_6} \cdot B_{\alpha_6}$ and $A_{\alpha_7} \cdot B_{\alpha_7}$, which causes the computation load of $L/3$ in Layer $2$ for each server. 

In this example, there are two recovery thresholds $R_1=5$ and $R_2=3$, corresponding to two layers, respectively. We term the choice of per-layer recovery thresholds as \emph{recovery profile}. There are totally $\widetilde{\gamma} = \gamma_1 + \gamma_2 =3$ coded matrices in a share where $\gamma_1$ coded matrices correspond to Layer $1$ and $\gamma_2$ coded matrices correspond to Layer $2$. 
Specifically, $\forall i \in [N]$, the shares $\widetilde{\mathcal{A}}_{i}$ and $\widetilde{\mathcal{B}}_{i}$ contain
\begin{align}
\widetilde{{A}}_{i,1} = A_{\alpha_i}, ~~ \widetilde{{A}}_{i,2} =  A_{\alpha_6,1}+ \alpha_i A_{\alpha_6,2}, ~~\widetilde{{A}}_{i,3} = A_{\alpha_7,1}+ \alpha_i A_{\alpha_7,2}, \\
	\widetilde{{B}}_{i,1} =B_{\alpha_i},~~ \widetilde{{B}}_{i,2} =  B_{\alpha_6,1}+ \alpha_i B_{\alpha_6,2}, ~~\widetilde{{B}}_{i,3} =B_{\alpha_7,1}+ \alpha_i B_{\alpha_7,2},
\end{align} 
respectively.
Each server needs to store all the above 6 coded matrices under the one-round communication model, but only 2 coded matrices at a time under the multi-round communication. Each server computes up to $\widetilde{\gamma} =3$ tasks in order, independent of the progress of the other servers. 
\end{exa}
\begin{table*}[!ht]
\centering
\caption{Calculation tasks in each server for Example 1.}\label{table: example 1}
\begin{tabular}{|c|c|c|c|c|c|}
	
	\hline
	&Server $1$ & Server $2$ & Server $3$ & Server $4$ & Server $5$ \\
	\hline
	Layer $1$ &
	$A_{\alpha_1} \cdot B_{\alpha_1}$ &
	$A_{\alpha_2} \cdot B_{\alpha_2}$ &
	$A_{\alpha_3} \cdot B_{\alpha_3}$ &
	$A_{\alpha_4} \cdot B_{\alpha_4}$ &
	$A_{\alpha_5} \cdot B_{\alpha_5}$ \\
	\hline
	Layer $2$ &
	\makecell[c]{$(A_{\alpha_6,1}+ \alpha_1 A_{\alpha_6,2})$ \\ $\cdot (\alpha_1 B_{\alpha_6,1}+  B_{\alpha_6,2})$,\\
		$(A_{\alpha_7,1}+ \alpha_1 A_{\alpha_7,2})$ \\ $\cdot (\alpha_1 B_{\alpha_7,1}+  B_{\alpha_7,2})$} &
	\makecell[c]{$(A_{\alpha_6,1}+ \alpha_2 A_{\alpha_6,2})$ \\ $\cdot (\alpha_2 B_{\alpha_6,1}+  B_{\alpha_6,2})$,\\
		$(A_{\alpha_7,1}+ \alpha_2 A_{\alpha_7,2})$ \\ $\cdot (\alpha_2 B_{\alpha_7,1}+  B_{\alpha_7,2})$} &
	\makecell[c]{$(A_{\alpha_6,1}+ \alpha_3 A_{\alpha_6,2})$ \\ $\cdot (\alpha_3 B_{\alpha_6,1}+  B_{\alpha_6,2})$,\\
		$(A_{\alpha_7,1}+ \alpha_3 A_{\alpha_7,2})$ \\ $\cdot (\alpha_3 B_{\alpha_7,1}+  B_{\alpha_7,2})$} &
	\makecell[c]{$(A_{\alpha_6,1}+ \alpha_4 A_{\alpha_6,2})$ \\ $\cdot (\alpha_4 B_{\alpha_6,1}+  B_{\alpha_6,2})$,\\
		$(A_{\alpha_7,1}+ \alpha_4 A_{\alpha_7,2})$ \\ $\cdot (\alpha_4 B_{\alpha_7,1}+  B_{\alpha_7,2})$} &
	\makecell[c]{$(A_{\alpha_6,1}+ \alpha_5 A_{\alpha_6,2})$ \\ $\cdot (\alpha_5 B_{\alpha_6,1}+  B_{\alpha_6,2})$,\\
		$(A_{\alpha_7,1}+ \alpha_5 A_{\alpha_7,2})$ \\ $\cdot (\alpha_5 B_{\alpha_7,1}+  B_{\alpha_7,2})$} \\
	\hline
	
\end{tabular}
\vspace{-0.5cm}
\end{table*}

\begin{figure}[!h]
\centering
\includegraphics[width=0.45\textwidth]{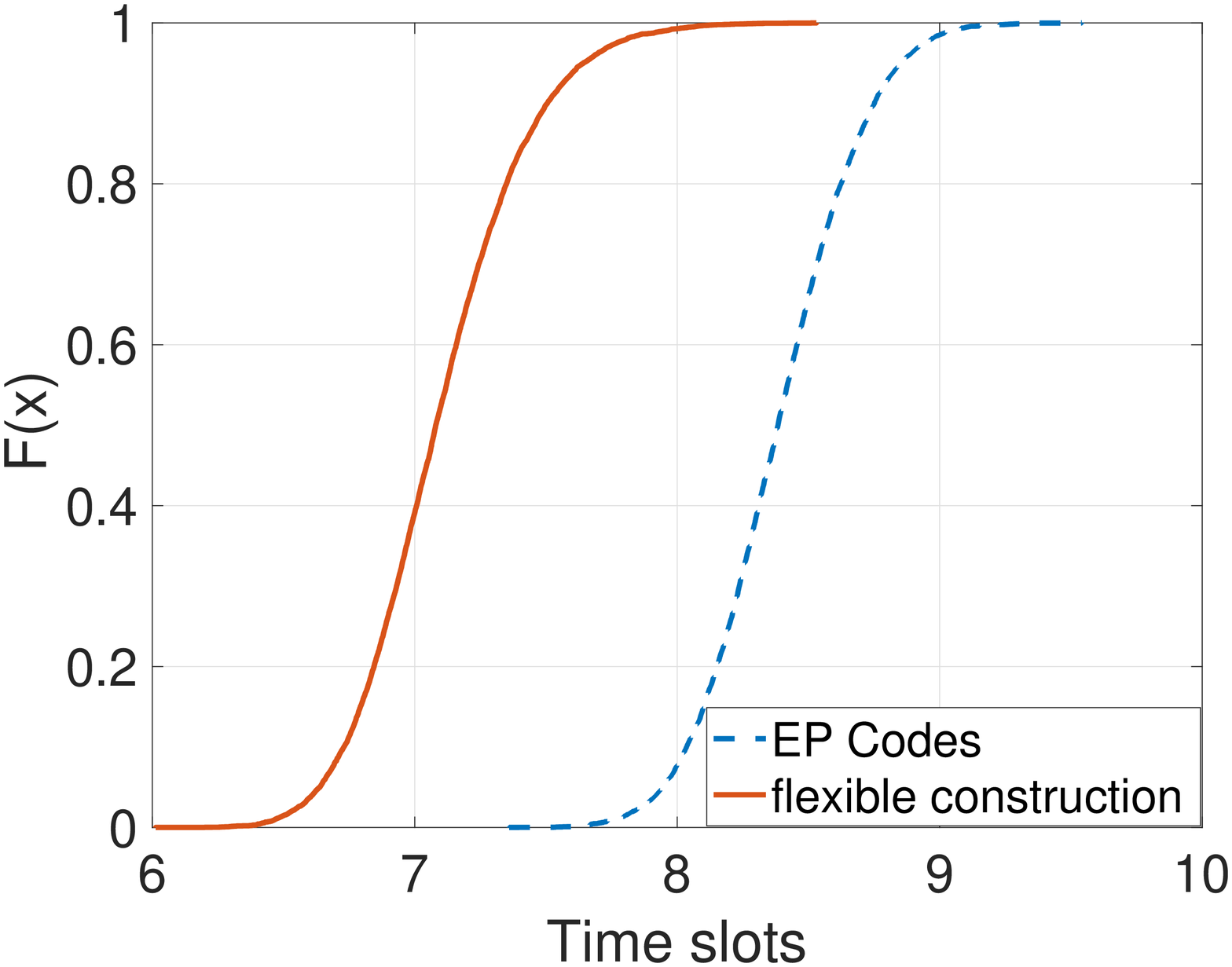}
\vspace{-0.5cm}
\caption{CDF of computation latency for flexible construction and EP code in Example 1 of Section \ref{construction}. $N=R_1=5,R_2=R=3$. We assume $\lambda = \kappa = \mu = 6U$, for some integer $U$, and the computation delay for multiplication of two $U \times U$ matrices in each server satisfy the exponential distribution with parameter $0.1$. The latency of the EP code is the delay of the $3$rd quickest server, and the slowest $2$ servers are viewed as stragglers. For the flexible construction, the computation is completed in the cases of $5$ servers complete $1$ task (no straggler), or $4$ servers complete  $2$ tasks (1 straggler), or $3$ servers complete $3$ tasks (2 stragglers). The overall latency is the smallest latency of these $3$ cases. The expected latency is $10.79$ for EP code, and $8.20$ for the flexible construction. Hence we save $24\%$. 
}\label{CDF1}
\vspace{-0.5cm}
\end{figure}

For Example \ref{example 1}, the computation load of each server is $L/3, L/2, 2L/3$ for the cases of no stragglers, $1$ straggler and $2$ stragglers, respectively. When there is no straggler (which is more likely in most practical systems), the computation load of each server is reduced $33\%$, from $L/2$ to $L/3$. 
The resulting computation latency under an exponential model is plotted in Fig. \ref{CDF1}.

In this example, if there is only one communication round from the sources to the servers, the storage size required for each server  is $\frac{2\lambda\kappa}{3} + \frac{2\kappa\mu}{3}$ for our flexible construction and $\frac{\lambda\kappa}{2} + \frac{\kappa\mu}{2}$ for the EP code. We will discuss how to partition the matrices to obtain an advantageous computation load while maintaining the same storage size in Section \ref{optimization}.

Next, we present the general definitions and constructions of our flexible schemes. The key component is to generate extra parities during the encoding in each layer that will correspond to extra tasks to be completed by higher layers to compensate for more stragglers. 

Define the \emph{recovery profile} as a tuple of integers $(R_1, R_2, \cdots, R_a)$, where $N \ge R_1 > R_2 > ... > R_a=R$ and $a$ is some integer termed the \emph{number of layers}. Denote 
\begin{align}\label{number of packages}
\gamma_j =
\begin{cases}
1, &j = 1,\\
 (R_{j-1} - R_j)\sum\limits_{J=1}^{j-1}\gamma_J,& 2 \le j \le a,
\end{cases}
\end{align}
which will be shown to be the number of tasks in each layer.
For two matrices $\Phi, \Psi$ and partition parameters $p_j,m_j,n_j$, define functions $f_j, g_j, j \in [a],$ as
\begin{align}\label{multiple-layer f}
	f_{j}(\alpha_i;\Phi) = \sum_{u=1}^{m_j}\sum_{v=1}^{p_j}{\Phi_{(u,v)}}\alpha_i^{v-1+p_j(u-1)},
\end{align}
\begin{align}\label{multiple-layer g}
	g_{j}(\alpha_i;\Psi) =  \sum_{u=1}^{p_j}\sum_{v=1}^{n_j}{\Psi_{(u,v)}}\alpha_i^{p_j-u+p_jm_j(v-1)},
\end{align}
where 
\begin{align}\label{eq:partition}
	\Phi = \left[\begin{array}{c c c} 
		\Phi_{(1,1)} & \cdots & \Phi_{(1,p_j)}  \\
		\Phi_{(2,1)} & \cdots & \Phi_{(2,p_j)}  \\
		\vdots  & \vdots &  \vdots \\
		\Phi_{(m_j,1)}  & \cdots & \Phi_{(m_j,p_j)}  \\
	\end{array}\right], 
	\Psi = \left[\begin{array}{c c c} 
		\Psi_{(1,1)}   & \cdots & \Psi_{(1,n_j)}  \\
		\Psi_{(2,1)}  & \cdots & \Psi_{(2,n_j)}  \\
		\vdots  & \vdots &  \vdots \\
		\Psi_{(p_j,1)}  & \cdots & \Psi_{(p_j,n_j)}  \\
	\end{array}\right].
\end{align} 
Note that \eqref{multiple-layer f} and \eqref{multiple-layer g} are the encoding functions of the EP codes \cite{Yu_Maddah-Ali_Avestimehr} used in Layer $j$.

\begin{cnstr}\label{multiple-layer construction}
Given recovery profile $(R_1, R_2, \cdots, R_a)$ and partitioning parameters $p_j, m_j, n_j$ such that $R_j = p_jm_jn_j + p_j - 1, j\in[a]$, the construction consists of $a$ layers. Fix $N+R_1-R_a$ distinct elements $\alpha_i, i\in[N+R_1-R_a],$ in a finite field $\mathbb{F}$.

In Layer $1$, set $A^{(1,1)} = A$ and $B^{(1,1)} = B$. A pair of coded matrices $f_{1}\left(\alpha_{t};A^{(1,1)}\right)$ and $g_{1}\left(\alpha_{t};B^{(1,1)}\right)$ are generated for Server $t$, $t\in[N]$. Moreover, extra $R_1-R_a$ pairs of parities will be generated, i.e., $f_{1}\left(\alpha_{N+t};A^{(1,1)}\right)$ and $g_{1}\left(\alpha_{N+t};B^{(1,1)}\right)$, $t\in[R_1-R_a]$. They will be used in higher layers. 

In Layer $j,2 \leq j \leq a$, the number of pairs of coded matrices is $\gamma_j$ given by \eqref{number of packages}.
For each $\delta_j \in [\gamma_j]$, a pair of coded matrices $f_{j}\left(\alpha_{t};A^{(j,\delta_j)}\right)$ and $g_{j}\left(\alpha_{t};B^{(j,\delta_j)}\right)$ are generated for Server $t$, $t\in[N]$. Besides, extra parities $f_j(\alpha_{N+t};A^{(j,\delta_j)})$ and $f_j(\alpha_{N+t};B^{(j,\delta_j)})$, $t \in [R_j-R_a]$, are produced for higher layers. 
Here, $A^{(j,\delta_j)}$ and $B^{(j,\delta_j)}$, $\delta_j \in [\gamma_j]$,  are from the extra parities $f_{J}(\alpha_{N+t};A^{(J,\delta_J)}),  g_{J}(\alpha_{N+t};B^{(J,\delta_J)})$ in Layer $J$ for all $J \in [j-1]$ and 
\begin{align}\label{eq:t_delta}
    R_j-R_{a}+1 \le t \le R_{j-1}-R_a,\delta_J\in[\gamma_J]. 
\end{align}
Specifically, given $j$ and $\delta_j$, $A^{(j,\delta_j)}$ and $B^{(j,\delta_j)}$ are set as
\begin{align}
	A^{(j,\delta_j)} = f_{J}\left(\alpha_{N+t}:A^{(J,\delta_J)}\right),\label{eq:encoding 1} \\
	B^{(j,\delta_j)} = g_{J}\left(\alpha_{N+t}:B^{(J,\delta_J)}\right), \label{eq:encoding 2} 
\end{align}
where
\begin{align}\label{eq:t}
t= \left\lfloor \delta_j  \frac{R_{j-1} - R_j}{\gamma_j} \right\rfloor + R_{j}-R_{a}+1,
\end{align}
and $J$ is the integer satisfying
\begin{align}\label{eq:J}
\sum_{x=1}^{J-1}\gamma_x < \delta_j \text{ mod } \frac{\gamma_j}{(R_{j-1} - R_j)} \leq  \sum_{x=1}^{J}\gamma_x
\end{align}
and
\begin{align}\label{eq:delta}
\delta_J= \delta_j \text{ mod } \frac{\gamma_j}{(R_{j-1} - R_j)} - \sum_{x=1}^{J-1}\gamma_x.
\end{align}
Intuitively, the $t$-th extra parities in all previous layers are encoded in Layer $j$, for all $t$ satisfying \eqref{eq:t_delta}. Equations \eqref{eq:t}, \eqref{eq:J}, and \eqref{eq:delta} simply mean that these extra parities are ordered from left to right and from top to bottom (see Fig. \ref{exa: 3 Layers} for an example).

Denote $\Gamma_{j,\delta_j}(\alpha_i)$ as the $\delta_j$-th task in Layer $j$ calculated in Server $i$, for $i\in[N], j\in[a],\delta_j\in[\gamma_j]$, where
\begin{align}\label{multiple-layer gamma}
	\Gamma_{j,\delta_j}(\alpha_i) = f_{j}\left(\alpha_i;A^{(j,\delta_j)}\right) \cdot  g_{j}\left(\alpha_i;B^{(j,\delta_j)}\right),
\end{align}
The calculation tasks of the construction are shown in Table \ref{multiple-layer}.

Note that there are in total $\widetilde{\gamma} = \sum_{j=1}^{a}\gamma_j$ tasks. The shares and the tasks are
\begin{align}
&\widetilde{\mathcal{A}}_{i} = \{ A^{(j,\delta_j)} \mid  j\in[a], \delta_j \in[\gamma_j] \}, \\
&\widetilde{\mathcal{B}}_{i} = \{ B^{(j,\delta_j)} \mid  j\in[a], \delta_j \in[\gamma_j] \}, \\
&\widetilde{{\Gamma}}_{i,  \sum\limits_{x=1}^{j-1}\gamma_x+\delta_j} = \Gamma_{j,\delta_j}(\alpha_i).
\end{align}

\end{cnstr}

\begin{table}[!ht]
\centering
\caption{Calculation tasks in each server for the multiple-layer construction, where $\delta_j$ ranges between $1$ and $\gamma_j$ as defined in \eqref{number of packages}, $j\in[a]$. }\label{multiple-layer}
\begin{tabular}{|c|c|c|c|c|c|c|c|}
	
	\hline
	&Server $1$ &  $\dots$ & Server $N$ & Extra parity $1$ & $\dots$ & $\dots$ & Extra parity $R_1-R_a$\\
	\hline
	Layer $1$ &
	$\Gamma_{1,1}(\alpha_1)$ &
	$\dots$ &
	$\Gamma_{1,1}(\alpha_N)$ &
	$\Gamma_{1,1}(\alpha_{N+1})$ &
	$\dots$ &
	$\dots$ &
	$\Gamma_{1,1}(\alpha_{N+R_1-R_a})$ \\
	\hline
	Layer $2$ &
	$\Gamma_{2,\delta_2}(\alpha_1)$ &
	$\dots$ &
	$\Gamma_{2,\delta_2}(\alpha_N)$ &
	$\Gamma_{2,\delta_2}(\alpha_{N+1})$ &
	$\dots$ &
	$\Gamma_{2,\delta_2}(\alpha_{N+R_2 - R_a})$ &
	\multicolumn{1}{>{\columncolor{mygray}}l |}{  }  \\
	\hline
	$\vdots$ &
	$\vdots$ &
	$\ddots$ &
	$\vdots$ &
	$\vdots$ & 
	$\vdots$ &
	\multicolumn{1}{>{\columncolor{mygray}}l |}{  } &
	\multicolumn{1}{>{\columncolor{mygray}}l |}{  } \\
	\hline
	Layer $a$ &
	$\Gamma_{a,\delta_a}(\alpha_1)$ &
	$\dots$ &
	$\Gamma_{a,\delta_a}(\alpha_N)$ &
	\multicolumn{1}{>{\columncolor{mygray}}l |}{  } &
	\multicolumn{1}{>{\columncolor{mygray}}l |}{  } &
	\multicolumn{1}{>{\columncolor{mygray}}l |}{  } &
	\multicolumn{1}{>{\columncolor{mygray}}l |}{  }\\
	\hline
\end{tabular}
\vspace{-0.5cm}
\end{table}

\begin{exa}
An example of a $3$-layer construction is shown in Fig. \ref{exa: 3 Layers}. We set $N=5, R = 2, (R_1, R_2, R_3) = (5, 3, 2)$. In Fig. \ref{exa: 3 Layers}, we show that the coded matrices transmitted from Source $1$ and Source $2$ are similar. In Layer $1$ ($A^{(1,1)} = A$), the coded matrices $f_{1}(\alpha_i;A^{(1,1)})$ are transmitted to Server $i$, $i\in[5]$, and $f_{1}(\alpha_{5+t};A^{(1,1)}), t\in[3]$ are the extra parities. These parities are used in Layers 2 and 3.  Specifically, $A^{(2,1)} = f_{1}(\alpha_{7};A^{(1,1)})$ and $A^{(2,2)} = f_{1}(\alpha_{8};A^{(1,1)})$ are used in Layer $2$ and $A^{(3,1)} = f_{1}(\alpha_{6};A^{(1,1)})$ is used in Layer $3$.
In Layer $2$, $f_{2}(\alpha_i;A^{(2,\delta_2)}),\delta_2 \in [2], i\in[5]$ are encoded using the above extra parities from Layer $1$. The generated extra parities $A^{(3,2)} = f_{2}(\alpha_{6};A^{(2,1)})$ and $A^{(3,3)} = f_{2}(\alpha_{6};A^{(2,2)})$ are used in Layer $3$.

\begin{figure}[!h]
		\centering
		\includegraphics[width=\textwidth]{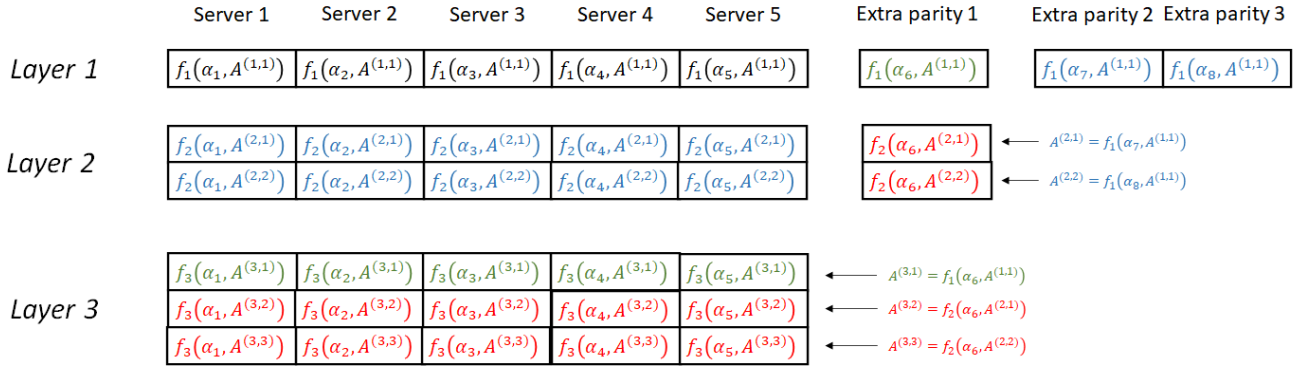}
		\caption{Example of coded matrices for $3$-layer construction, $N=R_1=5,R_2=3,R_3=R=2$.}\label{exa: 3 Layers}
\end{figure}	
\end{exa}

Theorem \ref{thm: Construction 1}, below, states the performance of the flexible construction in terms of storage and computation. This result  is based on the following decoding strategy: in the presence of $\hat{R}$ available servers, $R_j \le \hat{R} < R_{j-1}$, all tasks in Layers $1,2,\dots,j-1$ and some tasks in Layer $j$ are executed. It should be noted that the sum of storage sizes in all layers corresponds to the one-round communication model. However, under the multi-round communication model, the server storage size is only the maximum over the pairs of coded matrices. Since the coded matrix in a layer is encoded from sub-matrices in the previous layer, the higher the layer is, the smaller the size becomes. Hence, the storage size is just that of the first pair of coded matrices.
	
\begin{thm}\label{thm: Construction 1}
In Construction \ref{multiple-layer construction}, assume we have $\hat{R}$ available servers and $R\leq \hat{R} \leq N$, we need
\begin{align}\label{eq:general L_flex}
L_{\text{flex}} = 
\begin{cases}
	L_1, & \hat{R} \geq R_1,\\
	\left(1+\frac{R_{j-1} -\hat{R}}{p_jm_jn_j}\right)\sum\limits_{J=1}^{j-1} L_{J}, &R_{j} \leq \hat{R} < R_{j-1}, j \ge 2,
\end{cases} 
\end{align} 
computation load at each server to obtain the final result, where 
\begin{align}\label{eq:general L_j}
L_{j} = 
\begin{cases}
	\frac{\lambda\kappa\mu}{m_1 p_1 n_1},& j = 1,\\
	\frac{R_{j-1} - R_{j}}{p_jm_jn_j} \sum\limits_{J=1}^{j-1} L_{J},& j\ge 2,
\end{cases}
\end{align} 
is the total computation load at each server in Layer $j$.
The server storage size required in Layer $j$ is
\begin{align}\label{eq: storage in Layer j 1}
	C_j =  C_{j,A} +  C_{j,B},
\end{align}
where
\begin{align} \label{eq: storage in Layer j 2} 
C_{j,A} = 
	\begin{cases}
	\frac{\lambda\kappa}{p_1m_1}, & j=1, \\
	\frac{R_{j-1} - R_{j}}{p_j m_j} \sum\limits_{J=1}^{j-1} C_{J,A},& j\ge 2.
	\end{cases}
\end{align}
\begin{align}\label{eq: storage in Layer j 3}
C_{j,B} = 
	\begin{cases}
	\frac{\kappa\mu}{p_1n_1}, & j=1, \\
\frac{R_{j-1} - R_{j}}{p_j n_j} \sum\limits_{J=1}^{j-1} C_{J,B},& j\ge 2.
	\end{cases}
\end{align}
\end{thm}

\begin{IEEEproof}
In the following, we first prove \eqref{eq:general L_j}. Then, we show that with the computation load in \eqref{eq:general L_flex}, the central node is able to obtain the matrix product. At last, we prove the storage size required in each layer.

In Layer $j=1$, from \eqref{eq:partition}, we know that $f_{1}(\alpha_i;A^{(1,1)})$ and $f_{1}(\alpha_i;B^{(1,1)})$ have sizes $\frac{\lambda}{m_1} \times \frac{\kappa}{p_1}$ and $\frac{\kappa}{p_1} \times \frac{\mu}{n_1}$, respectively. Thus, the computation load in Layer $1$ is
\begin{align}
	L_1 = \frac{\lambda\kappa\mu}{m_1 p_1 n_1}.
\end{align}
In Layer $j$,  according to \eqref{eq:partition}, \eqref{multiple-layer f}, \eqref{multiple-layer g}, and \eqref{multiple-layer gamma}, the computation load of $\{\Gamma_{j,\delta_j}(\alpha_i) = f_{j}(\alpha_{i};A^{(j,\delta_j)}) \cdot g_{j}(\alpha_{i};B^{(j,\delta_j)}): \delta_j \in [\gamma_j]\}$ is $1/(p_jm_jn_j)$ fraction of that of $\mathcal{X} \triangleq \{A^{(j,\delta_j)}\cdot B^{(j,\delta_j)}: \delta_j \in[\gamma_j]\}$. Moreover, the computation load of $\mathcal{Y}(J,t) \triangleq \{f_{J}(\alpha_{N+t}:A^{(J,\delta_J)})\cdot f_{J}(\alpha_{N+t}:B^{(J,\delta_J)}):\delta_J\in[\gamma_J]\}$ is equal to the load (per server) at the $J$-th layer, which is $L_J$. Based on \eqref{eq:t_delta}, \eqref{eq:encoding 1} and \eqref{eq:encoding 2}, the computation load of $\mathcal{X}$ is equal to the load of $\mathcal{Y}(J,t)$ for all $J \in [j-1], R_j-R_a+1 \le t \le R_{j-1}-R_a$,
which is $(R_{j-1}-R_j)\sum\limits_{J=1}^{j-1} L_{J}$. Therefore, \eqref{eq:general L_j} is satisfied.

In the case that the number of available servers $\hat{R} \geq R_1$, according to the correctness of EP codes \cite{Yu_Maddah-Ali_Avestimehr}, the required results can be obtained by collecting $R_1$ evaluation points of $\Gamma_{1,1}(\alpha_i) $. Thus, we only need the computation in Layer $1$. 
		
In the case that $R_{j} \leq \hat{R} < R_{j-1}$, we first calculate all the tasks in Layers $1$ to $ j-1$, whose computation load is $\sum\limits_{J=1}^{j-1} L_{J}$. Then, in Layer $j$, Server $i$ calculates $\frac{R_{j-1}-\hat{R}}{R_{j-1}-R_j}\gamma_j$ tasks, i.e., $\Gamma_{j,\delta_j}(\alpha_{i}),\delta_j = 1,2,...,\frac{R_{j-1}-\hat{R}}{R_{j-1}-R_j}\gamma_j, i\in[N]$. Thus, the total computation is $\left(1+\frac{R_{j-1} -\hat{R}}{p_jm_jn_j}\right)\sum\limits_{J=1}^{j-1} L_{J}$.

{\bf Claim:} $R_J$ evaluations from $\{\Gamma_{J,\delta_J}(\alpha_i), i \in [N]\}$ can be obtained by the above calculations for $J=j,j-1,\dots,2,1$. 

We prove it by induction on $J$. As a consequence, the polynomial $\Gamma_{J,\delta_J}(\cdot)$ is decoded due to the correctness of EP codes \cite{Yu_Maddah-Ali_Avestimehr}. Hence, the final result can be decoded with $J=1$ and \eqref{eq:general L_flex} is proved.		

\textbf{Base case}:
In Layer $j$, since $\hat{R} \ge R_j$, the claim holds trivially.

		
\textbf{Induction step}: Suppose the claim holds for Layers $j,j-1,\dots,J+1$. We show that it will hold for Layer $J$. Note that $J<j$. The associated polynomials are decoded in Layers $j,j-1,\dots,J+1$. Then, from Eqs. \eqref{eq:t_delta}, \eqref{eq:encoding 1} and \eqref{eq:encoding 2}, one can calculate 
$\Gamma_{J,\delta_J}(\alpha_{N+t}),$ for $R_{j}-R_a+1 \le t \leq R_{j-1} -R_a - (\hat{R}-R_{j})$ from Layer $j$ and $R_{J'}-R_a+1 \le t \leq R_{J'-1} -R_a$ from Layers $J'=j-1,\dots,J+1$.
In total, $R_{J} - \hat{R}$ extra parities are obtained for the polynomial $\Gamma_{J,\delta_J}(\cdot)$.
Thus, together with $\hat{R}$ available nodes, $R_J$ evaluation points of $\Gamma_{J,\delta_J}(\cdot)$ are obtained, for all $\delta_J\in[\gamma_J]$.
		
		
The proof of the storage size is similar to the proof of \eqref{eq:general L_j}. The  proof sketch is as follows.

In Layer $1$, the server needs to store $f_{1}(\alpha_i;A^{(1,1)}), f_{1}(\alpha_i;B^{(1,1)}), i\in[N]$, then 
\begin{align}\label{multiple-layer storage 1}
C_1 = \frac{1}{p_1} \left( \frac{\lambda\kappa}{m_1} + \frac{\kappa\mu}{n_1} \right).
\end{align}
		
In Layer $j\ge 2$, from \eqref{number of packages}, \eqref{eq:encoding 1} and \eqref{eq:encoding 2},  the $\gamma_j$ tasks in Layer $j$ are encoded from the extra parities in Layers $1$ to $j-1$. Based on \eqref{eq:partition}, \eqref{multiple-layer f}, \eqref{multiple-layer g}, and \eqref{multiple-layer gamma}, the size of $f_{j}(\alpha_i;A^{(j,\delta_j)}), i\in[N]$ is $p_jm_j$ fractions of $A^{(j,\delta_j)}$, and the size of $f_{j}(\alpha_i;B^{(j,\delta_j)}), i\in[N]$ is $p_jn_j$ fractions of $B^{(j,\delta_j)}$. Thus, \eqref{eq: storage in Layer j 2} and \eqref{eq: storage in Layer j 3} are obtained.	
\end{IEEEproof}

\begin{rem}
In Fig. \ref{exa: 3 Layers}, partial computation results can be also utilized to accelerate the computation in several cases such that the nodes contribute different number of results depending on their speed. For example, when Servers $1$ and $2$ complete their first $4$ tasks and Server $3$ completes its first $2$ tasks, we are able to obtain $f_1(\alpha_i,A^{(1,1)})$ for $i=1,2,3,6,7$, thus obtain the final results. Similar partial results utilization can be found in our general constructions, but in this paper we assume a server is either available or not able to provide any results.
\end{rem}

\begin{rem}
CSA codes and GCSA codes \cite{Jia_Jafar_CDBC} are designed to handle batch processing of matrix multiplication, namely, the multiplication of two sequences of matrices. They also provide solutions for secure distributed computation. Combined with these codes, our construction can be easily modified to handle batch processing and secure distributed computation.
\end{rem}

The following corollary states a special case of the computation load that will be useful in the optimization discussed in Section \ref{optimization} under the multi-round communication model.
	
\begin{cor}\label{cor:optimal case}
	In the case of $p_j = 1, j\ge2$, we have $m_jn_j = R_j$ in Construction \ref{multiple-layer construction}. The $j$-th layer's computation load of each server is
	\begin{align}\label{eq:special L_j}
		L_j = 			
		\begin{cases}
			\frac{\lambda\kappa\mu}{m_1 p_1 n_1}, & j = 1,\\
			\frac{R_1(R_{j-1}-R_j)}{R_{j-1}R_j}L_1, & j \ge 2,
		\end{cases} 
	\end{align}
and the total computation of each server is
	\begin{align}\label{eq:computation load in model 2}
		L_{\text{flex}} =
		\begin{cases}
			L_1, & R_{1} \leq \hat{R} ,\\
			\frac{R_1(R_j+R_{j-1}-\hat{R})}{R_{j-1}R_j} L_1, &R_{j} \leq \hat{R} < R_{j-1}, j \ge 2,
		\end{cases} 
	\end{align}
where $\hat{R}$ is the number of non-straggler servers. Specifically, when $\hat{R} = R_j$,
	\begin{align}
		L_{\text{flex}} = \frac{R_1}{R_j}L_1. 
	\end{align}
\end{cor}
	
\begin{IEEEproof}
	We first prove \eqref{eq:special L_j} by induction.
	
	\textbf{Base case}: When $j=2$, we get $L_2 = \frac{R_{1}-R_2}{R_2}L_1$ from \eqref{eq:general L_j} and it satisfies \eqref{eq:special L_j}.

	\textbf{Induction step}: Suppose $L_2, ... , L_j$ satisfy \eqref{eq:special L_j}. From \eqref{eq:general L_j} and $p_jm_jn_j = R_j$ we know that
	\begin{align}
		L_j = \frac{R_{j-1} - R_{j}}{R_j}\sum_{J=1}^{j-1}L_J . 
	\end{align}
	Then, we have
	\begin{align}
		L_{j+1} &=  \frac{R_{j} - R_{j+1}}{R_{j+1}}\sum_{J=1}^{j-1}L_J + \frac{R_{j} - R_{j+1}}{R_{j+1}}L_j \\ \nonumber
		& = \frac{R_{j} - R_{j+1}}{R_{j+1}}\frac{R_j}{R_{j-1}-R_j}L_j + \frac{R_{j} - R_{j+1}}{R_{j+1}}L_j  \\ \nonumber
		& = \frac{R_{j-1}(R_j-R_{j+1})}{R_{j+1}(R_{j-1}-R_j)}L_j \\ \nonumber
		& = \frac{R_1(R_{j}-R_{j+1})}{R_{j}R_{j+1}}L_1, 
	\end{align}
	which satisfies \eqref{eq:special L_j}.
	
	Then, for the total computation, from \eqref{eq:general L_flex} we can easily check that for $R_{j} \leq \hat{R} < R_{j-1}, j\in[a]$, we have
	\begin{align}
		L_{\text{flex}} &= \left(1+\frac{R_{j-1} -\hat{R}}{R_j}\right)\sum\limits_{J=1}^{j-1} L_{J} \\ \nonumber
		& = \left(1+\frac{R_{j-1} -\hat{R}}{R_j}\right)\frac{R_j}{R_{j-1}-R_j}L_j \\ \nonumber
		& = \frac{R_1(R_j+R_{j-1}-\hat{R})}{R_{j-1}R_j}L_1.
	\end{align}
	The proof is completed.
\end{IEEEproof}


\section{Computation Load Optimization} \label{optimization}
In this section, we discuss how to pick the matrix partition parameters and the recovery profile to optimize the computation load given the storage capacity. Under the one-round communication model, we find the optimal parameters for the 2-layer flexible construction and show that when the storage capacity is above a threshold, the flexible construction outperforms the fixed EP code. For the multi-round communication model, we show that all layers except the first layer reduce to block-wise matrix-vector multiplication, and when the straggler probability is small, the most number of layers is optimal.



Recall $\hat{R}$ is the number of non-straggler servers, $R \le \hat{R} \le N$. We consider the expected computation load over the realizations of $\hat{R}$.
Assume for each instance of computing, $\hat{R}$ is independent and identically distributed.
Denote $q_{j}$ as the probability of $j$ stragglers in the system. Formally,
\begin{align}
	&q_j = P(\hat{R} = N-j), ~~ \forall j \in \{0, ,1, \cdots, N-R\}.
\end{align}
Here, $R$ is chosen such that the probability of having more than $N-R$ stragglers is negligible. Therefore, $j$ is assumed to be in the range between $0$ and $N-R$, and
\begin{align}
    \sum_{j=0}^{N-R} q_j =1.
\end{align}
The expectation of the computation load is 
\begin{align}\label{eq:expected_L}
	E[L_{\text{flex}}] &= \sum_{j=0}^{N-R}q_j L_{\text{flex}}(\hat{R}=N-j),
\end{align}
where $L_{\text{flex}}(\hat{R})$ is the computation load for $\hat{R}$ non-straggler servers. The goal is to minimize $E[L_{\text{flex}}]$ over the partitioning parameters $p_j,m_j,n_j$, $j \in [a]$ and the recovery profile $\{R_1, \cdots, R_a\}$, given the recovery threshold $R_a=R$ and the storage constraint $C$ in each server. 
Although in practical systems $p_j,m_j,n_j, R_j, j \in [a]$ are required to be integers, in this section, we only assume them as real numbers to simplify the optimization analysis. To find an integer solution (not necessarily optimal), we pick the parameters close to the optimal real values that satisfy the recovery threshold and the storage constraint.

\subsection{Optimization on Entangled Polynomial codes}
As a warm-up, let us start with an EP code with a fixed recovery threshold $R$, which satisfies $R=m_0p_0n_0+p_0-1$ according to \cite{Yu_Maddah-Ali_Avestimehr}, for some undetermined partition parameters $p_0,m_0,n_0$. The computation load of EP codes remains the same if the number of stragglers is no greater than $N-R$. According to \cite{Yu_Maddah-Ali_Avestimehr}, the computation load and the required storage size are
\begin{align}
	L_{\text{EP}}(\hat{R}) = 
	\frac{\lambda\kappa\mu}{m_0 p_0 n_0}, \hat{R} \ge R, 
\end{align}
\begin{align}
	C_{\text{EP}} = \frac{1}{p_0} \left( \frac{\lambda\kappa}{m_0} + \frac{\kappa\mu}{n_0} \right).
\end{align}

Thus, the optimization problem can be  formulated as
\begin{equation}\label{optimal 1}
	\begin{aligned}
		\min_{p_0,m_0,n_0} \quad & L_{\text{EP}} = \frac{\lambda\kappa\mu}{m_0 p_0 n_0}, \\
		\textrm{s.t.} \quad & R = p_0m_0n_0 + p_0 -1 , \\
		& \frac{\lambda\kappa}{p_0m_0} + \frac{\kappa\mu}{p_0n_0} \leq C . \\
	\end{aligned}
\end{equation}

\begin{thm}\label{thm:optimazation 1}
	The solution of the EP code optimization problem in \eqref{optimal 1} is
\begin{align}\label{eq:optimal L_EP}
 L_{\text{EP}}^\ast = \frac{2C\lambda\kappa\mu}{C(R+1)+\sqrt{C^2(R+1)^2-16\lambda\kappa^2\mu}}
\end{align}	
with	
\begin{align}\label{eq:optimal_p0}
	p_0^\ast = \frac{1}{2}(R+1) - \frac{1}{2}\sqrt{(R+1)^2-16\frac{\lambda\kappa^2\mu}{C^2}},
\end{align}
and $m_0^\ast,n_0^\ast$ are given by $m_0^\ast n_0^\ast = \frac{R+1}{p_0^\ast} - 1$ and $\lambda\kappa n_0^\ast = \kappa\mu m_0^\ast$. 
\end{thm}
\begin{IEEEproof}
	Using the threshold constraint
	\begin{align}\label{eq:threshold_constraint}
		m_0n_0 = \frac{R+1}{p_0} - 1, 
	\end{align}
	we have
	$L_{\text{EP}} =  \frac{\lambda\kappa\mu}{R + 1 - p_0},$
	which is an increasing function of $p_0$.
	So, we minimize $p_0$ under the constraint that
	\begin{equation}\label{constraint simplification 1}
		\frac{\lambda\kappa n_0 + \kappa\mu m_0}{R+1-p_0} \leq C.
	\end{equation}
	Note that 
	\begin{align}\label{approximation 1}
		\lambda\kappa n_0 + \kappa\mu m_0 &\geq 2\sqrt{\lambda\kappa^2\mu m_0 n_0} = 2\sqrt{\lambda\kappa^2\mu  \frac{R+1-p_0}{p_0}} 
	\end{align}
	and it holds with equality if and only if $\lambda\kappa n_0 = \kappa\mu m_0$. 
	Combining \eqref{constraint simplification 1} with \eqref{approximation 1} results in
	\begin{align}\label{storage constraint}
		2\sqrt{\frac{\lambda\kappa^2\mu}{(R + 1 - p_0)p_0}} \leq C.
	\end{align}
	Note that $2\sqrt{\frac{\lambda\kappa^2\mu}{(R + 1 - p_0)p_0}}$ decreases with $p_0$ because the derivative
	\begin{align}\label{approximation 2}
		\frac{\text{d}\,(R + 1 - p_0)p_0}{\text{d}\, p_0} = R + 1 -2 p_0 = p_0m_0n_0-p_0 \ge 0.
	\end{align}
	Therefore, $L_{\text{EP}}$ reaches its optimal value when $\lambda\kappa n_0 = \kappa\mu m_0$ and \eqref{storage constraint} holds with equality, i.e., $p_0^\ast = \frac{1}{2}(R+1) - \frac{1}{2}\sqrt{(R+1)^2-16\frac{\lambda\kappa^2\mu}{C^2}}$. As a result, $m_0^\ast,n_0^\ast$ can be obtained by $m_0^\ast n_0^\ast = \frac{R+1}{p_0^\ast} - 1$ and $\lambda\kappa n_0^\ast = \kappa\mu m_0^\ast$. The optimal computation load is
	\begin{align}
	L_{EP}^\ast =  \frac{2C\lambda\kappa\mu}{C(R+1)+\sqrt{C^2(R+1)^2-16\lambda\kappa^2\mu}}.
	\end{align}
	The theorem is proved.
\end{IEEEproof}


\begin{rem}
In Theorem \ref{thm:optimazation 1}, the storage capacity is required to satisfy
\begin{align}\label{minimum storage 1}
C \ge \frac{4\kappa\sqrt{\lambda\mu}}{1+R}.
\end{align}
to have a valid $L_{\text{EP}}^\ast$ in \eqref{eq:optimal L_EP}. In addition, by combining \eqref{constraint simplification 1} and \eqref{approximation 1}, we obtain the minimum storage required as $2\sqrt{\frac{\lambda\kappa^2\mu}{(R + 1 - p_0)p_0}}$ in \eqref{storage constraint}. Since $p_0 = \frac{R+1}{m_0n_0+1}$ and $m_0,n_0$ are at least 1, we conclude that \eqref{minimum storage 1} is the minimum storage constraint requirement to use EP codes for  distributed matrix multiplication. 
\end{rem}

\subsection{Optimization for the one-round communication model}

Next, we consider the flexible constructions with the one-round communication model. In this model, all the tasks are sent to the server in one communication round. Thus, the sum of the task sizes should not exceed the storage constraint. Since the more layers, the larger the total size of the tasks is, only the 2-layer construction is considered. We first optimize the partition parameters with predetermined $R_1,R_2=R$. After that, $R_1$ is optimized. 

By the expression of the computation load in Theorem \ref{thm: Construction 1}, the expectation of the computation load in \eqref{eq:expected_L} becomes 
\begin{align}
E [L_{\text{flex}}] &= \sum\limits_{j=0}^{N-R_2} q_j\frac{\lambda\kappa\mu}{p_1m_1n_1} +  \sum\limits_{j=N-R_1+1}^{N-R_2} q_j \frac{\lambda\kappa\mu (R_1 + j - N)}{m_1m_2 p_1p_2 n_1n_2}. \label{overall latency for 2 Layers}
\end{align}
In practical systems, the probability of having many stragglers is usually small. For instance, less than 110 failures occur over a 3000-node production cluster of Facebook per day \cite{sathiamoorthy2013xoring}. So, we ignore the second term in \eqref{overall latency for 2 Layers} and use the approximation $L_{\text{flex}} = \frac{\lambda\kappa\mu}{p_1m_1n_1}$ in our optimization problem.
Combined with \eqref{eq: storage in Layer j 1}, the optimization problem can be formulated as
\begin{equation}\label{optimal 2}
	\begin{aligned}
		\min_{p_1,m_1,n_1,p_2,m_2,n_2}  & L_{\text{flex}} = \frac{\lambda\kappa\mu}{p_1m_1n_1}, \\
		\textrm{s.t.} \quad & R_1 = p_1m_1n_1 + p_1 -1 , \\
		&  R_2 = p_2m_2n_2 + p_2 -1 , \\
		&  \frac{1}{p_1} \left( \frac{\lambda\kappa}{m_1} + \frac{\kappa\mu}{n_1} \right) + \frac{ (R_1-R_2)}{p_1p_2} \left( \frac{\lambda\kappa}{m_1m_2} + \frac{\kappa\mu}{n_1n_2} \right)  \leq C .  \\
	\end{aligned}
\end{equation}
It should be noted that when $R_1 = R_2 = R$, the 2-layer flexible construction reduces to the fixed EP code and the optimal partition parameters remain the same.
	
\begin{thm}\label{thm:optimazation 2}
	Fix $R_1,R_2=R$. The solution of the 2-layer flexible construction optimization \eqref{optimal 2} is
\begin{align}\label{eq:optimal L_flex}
L_{\text{flex}}^\ast = \frac{2C(R_2+1)\lambda\kappa\mu}{C(R_1+1)(R_2+1)+\sqrt{C^2(R_1+1)^2(R_2+1)^2-16\lambda\kappa^2\mu(2R_1-R_2+1)^2}},
\end{align}
with
	\begin{align}\label{eq:optimal_p1}
		p_1^\ast =  \frac{1}{2}(R_1+1) - \frac{1}{2}\sqrt{(R_1+1)^2-\frac{16\lambda\kappa^2\mu(2R_1-R_2+1)^2}{C^2(R_2+1)^2}},
	\end{align}
	and $m_1^\ast,n_1^\ast$ are given by $m_1^\ast n_1^\ast = \frac{R_1+1}{p_1^\ast} - 1$ and $\lambda\kappa n_1^\ast = \kappa\mu m_1^\ast$, and $p_2^\ast = \frac{R_2+1}{2},m_2^\ast = 1, n_2^\ast = 1$.
\end{thm}
\begin{IEEEproof}
	Using $m_1n_1 = \frac{R_1+1}{p_1} - 1$, we have
	$L_{\text{flex}} =  \frac{\lambda\kappa\mu}{R_1 + 1 - p_1}$, which is an increasing function of $p_1$. Therefore, to maximize $L_{\text{flex}}$ we need to minimize $p_1$.
	
	Using $m_1n_1 = \frac{R_1+1}{p_1} - 1$ and $m_2n_2 = \frac{R_2+1}{p_2} - 1$, similar to \eqref{constraint simplification 1} and \eqref{approximation 1}, we have:
	\begin{align}
	    & C  \nonumber\\
	    \ge & \frac{1}{p_1} \left(\frac{\lambda\kappa}{m_1} + \frac{\kappa\mu}{n_1}\right) 
	    + \frac{ (R_1-R_2)}{p_1p_2} \left( \frac{\lambda\kappa}{m_1m_2} + \frac{\kappa\mu}{n_1n_2} \right) \label{eq:storage_1}\\
	    \ge & 2\sqrt{\frac{\lambda\kappa^2\mu}{(R_1 + 1 - p_1)p_1}}
		+ 2(R_1-R_2)\sqrt{\frac{\lambda\kappa^2\mu}{(R_1 + 1 - p_1)(R_2 + 1 - p_2)p_1p_2}}. \label{eq:storage_2}
	\end{align}
	Here \eqref{eq:storage_1} holds with equality when $\lambda\kappa n_1 = \kappa\mu m_1$ and $\lambda\kappa n_1 n_2 = \kappa\mu m_1 m_2$ or $n_2=m_2$.
	Similar to \eqref{approximation 2}, 
	it is easy to show that \eqref{eq:storage_2} is a decreasing function of $p_1$ and $p_2$. For any fixed $p_2$, to obtain the minimum $p_1$, we should set \eqref{eq:storage_2} equal to $C$. When \eqref{eq:storage_2} is fixed, $p_1$ is minimized when $p_2$ reaches its maximum because a bigger $p_2$ results in a smaller $p_1$ when \eqref{eq:storage_2} is equal to $C$.
	Noticing that $p_2 = \frac{R_2+1}{m_2n_2+1}$ and $m_2,n_2$ are at least 1, we set $p_2^\ast = \frac{R_2+1}{2},m_2^\ast = 1, n_2^\ast = 1$. The optimal $p_1^\ast$ and $L_{\text{flex}}^\ast$ are obtained accordingly.
\end{IEEEproof}

If $R_2$ is odd, then the choices of $p_2,m_2,n_2$ in the above theorem are the exact optimal integer parameters.

\begin{rem}\label{rem:minimum storage constraint for 2-layer}
In Theorem \ref{thm:optimazation 2}, the storage capacity is required to satisfy
\begin{align}\label{minimum storage 2}
C \ge \frac{4\kappa\sqrt{\lambda\mu}(2R_1-R_2+1)}{(1+R_2)(1+R_1)}
\end{align}
to have a valid $L_{\text{flex}}^\ast$ in \eqref{eq:optimal L_flex}. In addition, we obtain the minimum storage required in \eqref{eq:storage_2}. Since $p_1 = \frac{R_1+1}{m_1n_1+1}, p_2 = \frac{R_2+1}{m_2n_2+1}$, and $m_1,n_1,m_2,n_2$ are at least 1, we conclude that \eqref{minimum storage 2} is the minimum storage constraint requirement to use our 2-layer flexible codes for the distributed matrix multiplication. When $R_1 = R_2$, \eqref{minimum storage 2} is the same as \eqref{minimum storage 1}. 
\end{rem}

Next, we provide an example for the optimal integer solutions of the partition parameters.

\begin{exa}
	Assume there are $N=8$ servers and we need to tolerate $N-R=1$ straggler. $\lambda = \kappa = \mu$ and the storage size of each server is limited by $C=\frac{8}{7}\lambda\kappa$. Using the EP code, the optimal choice of $\{p_0,m_0,n_0\}$ is $\{1,1,7\}$, which results in a storage size of $\frac{8}{7}\lambda\kappa$ and a computation load per server of $\frac{1}{7}\lambda\kappa\mu=0.143\lambda\kappa\mu$. Using the $2$-layer flexible construction with $R_1 = 8$ and $R_2 = 7$, the optimal parameters are chosen as $p_1 = 1,m_1 = 2,n_1 = 4,p_2 = 4,m_2 = 1,n_2 = 1$, which cost a storage size of $\frac{15}{16}\lambda\kappa$ and a computation load of $\frac{1}{8}\lambda\kappa\mu$ when there is no straggler, with an additional computation load of $\frac{1}{32}\lambda\kappa\mu$ when there is one straggler. Assuming the probability of one straggler to be $10\%$, the average computation load is $0.128 \lambda\kappa\mu$. In this example, we save both storage size and average computation load while maintaining one straggler tolerance.
\end{exa}
	
Having found the best computation load for a fixed recovery profile as in Theorem \ref{thm:optimazation 1}, next, we discuss the optimization of the recovery profile. Given the straggler tolerance level $N-R_2$, we just need to optimize $R_1$, such that $R_2 \le R_1 \le N$.
	
\begin{thm}\label{thm:optimal R_1}
To minimize the 2-layer computation load $L_{\text{flex}}^\ast$ in \eqref{eq:optimal L_flex}, the optimal $R^\ast_1$ is
\begin{align}\label{eq:optimal R_1}
R^\ast_1 = 			
\begin{cases}
N, & C \ge \frac{8\kappa\sqrt{\lambda\mu}}{R_2+1},\\
\min\left(N,\frac{ C^2(R_2+1)^2 (R_2+3)+64\lambda\kappa^2\mu(R_2-1)}{2(64\lambda\kappa^2\mu- C^2(R_2+1)^2 )}\right), &\frac{8\kappa}{R_2+1}\sqrt{\frac{\lambda\mu}{3}}< C < \frac{8\kappa\sqrt{\lambda\mu}}{R_2+1}, \\
R_2, & C \le \frac{8\kappa}{R_2+1}\sqrt{\frac{\lambda\mu}{3}}.
\end{cases}     
\end{align}
\end{thm}

\begin{IEEEproof}
The optimal computation given $R_1$ is shown in \eqref{eq:optimal L_flex}. Since the numerator is a constant not related to $R_1$, we set $Y$ as the denominator and $L^\ast_{\text{flex}}$ has the minimum value when $Y$ reaches its maximum.
\begin{align}\label{eq:derivate of Y}
\frac{dY}{dR_1} = C (R_2+1) + \frac{C^2(R_2+1)^2(R_1+1)-32\lambda\kappa^2\mu(2R_1-R_2+1)}{\sqrt{C^2(R_1+1)^2(R_2+1)^2-16\lambda\kappa^2\mu(2R_1-R_2+1)^2}}.
\end{align}
Setting $\frac{dY}{dR_1} = 0$, we have
\begin{align}
\left(\frac{32\lambda\kappa^2\mu(2R_1-R_2+1)}{C(R_2+1)}\right)^2  =64\lambda\kappa^2\mu(2R_1-R_2+1)(R_1+1)  -16\lambda\kappa^2\mu(2R_1-R_2+1)^2.
\end{align}
Since $R_1 \ge R_2$, the term $\lambda\kappa^2\mu(2R_1-R_2+1)\neq 0$ can be cancelled and the solution to the above equation is
\begin{align}\label{eq:R_1 with zero derivate}
\hat{R_1} \triangleq \frac{ C^2(R_2+1)^2 (R_2+3)+64\lambda\kappa^2\mu(R_2-1)}{2(64\lambda\kappa^2\mu- C^2(R_2+1)^2 )}.
\end{align}
Let $X=C^2(R_1+1)^2(R_2+1)^2-16\lambda\kappa^2\mu(2R_1-R_2+1)^2$. We have $X \ge 0$ due to the minimum storage constraint in Remark \ref{rem:minimum storage constraint for 2-layer}.
We simplify \eqref{eq:derivate of Y} as
\begin{align}\label{eq:simplified dY/dR_1}
\frac{dY}{dR_1} = C (R_2+1) + \frac{\frac{dX}{dR_1}}{2\sqrt{X}},
\end{align}
where
\begin{align}
\frac{dX}{dR_1} = 2(C^2(R_2+1)^2 - 64\lambda\kappa^2\mu) R_1 + 2C^2(R_2+1)^2 + 64\lambda\kappa^2\mu (R_2-1)
\end{align}
is a linear function of $R_1$ and the constant term $2C^2(R_2+1)^2 + 64\lambda\kappa^2\mu (R_2-1)>0$ since $R_2$ is at least 1.

In the case of $C \ge \frac{8\kappa\sqrt{\lambda\mu}}{R_2+1}$, we have 
\begin{align}
C^2(R_2+1)^2 - 64\lambda\kappa^2\mu \ge 0 \Rightarrow \frac{dX}{dR_1} > 0 \Rightarrow \frac{dY}{dR_1} > 0.
\end{align}
Thus, we should pick $R^\ast_1 = N$. 

In the case of $C < \frac{8\kappa\sqrt{\lambda\mu}}{R_2+1}$, we get
\begin{align}\label{eq:d^2X/dR^2_1}
    C^2(R_2+1)^2 - 64\lambda\kappa^2\mu < 0 \Rightarrow \frac{dX}{dR_1} \text{ is a decreasing linear function}.
\end{align}
We discuss $\frac{dY}{dR_1}$ when $R_1$ varies between $(0,\frac{C^2(R_2+1)^2 + 32\lambda\kappa^2\mu}{64\lambda\kappa^2\mu-C^2(R_2+1)^2}]$ and $(\frac{C^2(R_2+1)^2 + 32\lambda\kappa^2\mu}{64\lambda\kappa^2\mu-C^2(R_2+1)^2}, +\infty)$ separately.
In the first region, we have
\begin{align}
R_1 \leq \frac{C^2(R_2+1)^2 + 32\lambda\kappa^2\mu}{64\lambda\kappa^2\mu-C^2(R_2+1)^2} \Rightarrow \frac{dX}{dR_1} \ge 0 \Rightarrow \frac{dY}{dR_1} > 0,    
\end{align}
$Y$ reach its maximum when $R_1 = \frac{C^2(R_2+1)^2 + 32\lambda\kappa^2\mu}{64\lambda\kappa^2\mu-C^2(R_2+1)^2}$. In the second region, we have
\begin{align}\label{eq:dX/dR_1}
R_1 > \frac{C^2(R_2+1)^2 + 32\lambda\kappa^2\mu}{64\lambda\kappa^2\mu-C^2(R_2+1)^2} \Rightarrow \frac{dX}{dR_1} < 0.    
\end{align}
Clearly, $\frac{\frac{dX}{dR_1}}{2\sqrt{X}}$ is a decreasing function of $R_1$, because $\frac{dX}{dR_1}$ is a negative decreasing function of $R_1$ by \eqref{eq:d^2X/dR^2_1} and \eqref{eq:dX/dR_1}, and $\sqrt{X}$ is a positive decreasing function of $R_1$ by \eqref{eq:dX/dR_1}.
Then, with \eqref{eq:simplified dY/dR_1} we can conclude that 
\begin{align}
\frac{d^2Y}{dR^2_1} = \frac{d\frac{\frac{dX}{dR_1}}{2\sqrt{X}}}{dR_1} < 0.
\end{align}
In addition, we know from \eqref{eq:R_1 with zero derivate} that $\hat{R_1}$ is located in $(\frac{C^2(R_2+1)^2 + 32\lambda\kappa^2\mu}{64\lambda\kappa^2\mu-C^2(R_2+1)^2}, +\infty)$ when $R_2\ge 2$, and hence is a local maximum.
Therefore, combining the $2$ ranges of $R_1$, we conclude that $Y$ reaches its maximum in \eqref{eq:R_1 with zero derivate}. Finally, the proof is completed considering the requirement that $R_2 \le R_1 \le N$, and the fact that $\hat{R_1} \ge R_2$ is satisfied when $C\ge \frac{8\kappa}{R_2+1}\sqrt{\frac{\lambda\mu}{3}}$. 
\end{IEEEproof}

\begin{cor}
The flexible construction with 2 layers is better than a fixed EP code in terms of the computation load when the storage constraint $C$ satisfy:
\begin{align}\label{eq:requirement for 2 Layer}
C>\frac{8\kappa}{R_2+1}\sqrt{\frac{\lambda\mu}{3}}.
\end{align}
\end{cor}
\begin{IEEEproof}
From Theorems \ref{thm:optimazation 1} and \ref{thm:optimazation 2} we have
\begin{align}
L^\ast_{\text{flex}}|_{R_1=R_2} = L^\ast_{\text{EP}}.
\end{align}
Also, it is easy to check that $R^\ast_1 > R_2$ in \eqref{eq:optimal R_1} when \eqref{eq:requirement for 2 Layer} is satisfied. Then, combining Theorem \ref{thm:optimal R_1} we conclude that
\begin{align}
L^\ast_{\text{flex}}|_{R_1=R^\ast_1} < L^\ast_{\text{flex}}|_{R_1=R_2} = L^\ast_{\text{EP}}.
\end{align}
The proof is completed.
\end{IEEEproof}

We summarize how to choose the optimal constructions in different situations in Table \ref{table: optimal choice of constructions}.
\begin{table}
\centering
\caption{Optimal choices of the flexible constructions given the number of servers $N$, the failure tolerance $N-R$ and the storage constraint $C$.}\label{table: optimal choice of constructions}
\begin{tabular}{|c|c|c|}
	
\hline
Storage constraint $C$	& Optimal constructions & Optimal matrix partition \\
\hline
$C < \frac{4\kappa\sqrt{\lambda\mu}}{1+R_2}$ & Not available & Not available \\
\hline
$\frac{4\kappa\sqrt{\lambda\mu}}{1+R_2} \le C \le \frac{8\kappa}{R_2+1}\sqrt{\frac{\lambda\mu}{3}}$ & Fixed EP codes & $p_0,m_0,n_0$ chosen in Theorem \ref{thm:optimazation 1} \\
\hline
$C > \frac{8\kappa}{R_2+1}\sqrt{\frac{\lambda\mu}{3}}$ & Flexible codes with $R_1$ chosen in Theorem \ref{thm:optimal R_1} & $p_j,m_j,n_j,j \in [2]$ chosen in Theorem \ref{thm:optimazation 2} \\
\hline

\end{tabular}
\end{table}

\begin{figure}
\centering
\includegraphics[width=.7\textwidth]{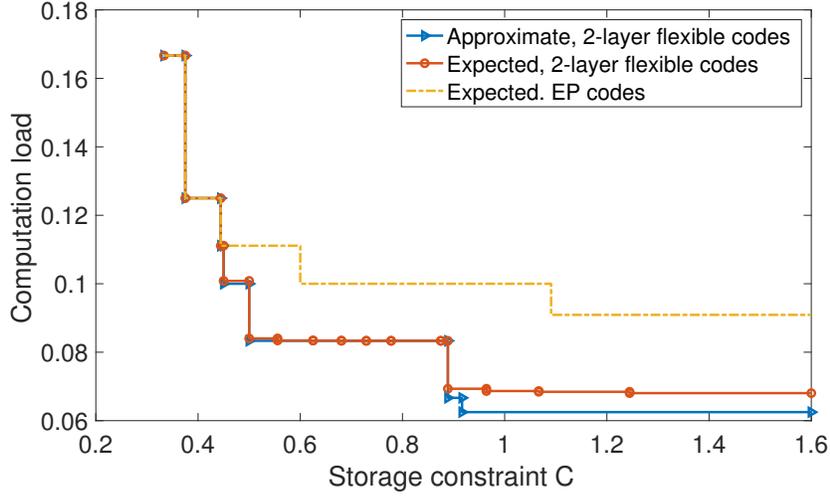}\caption{Computation load comparison of our 2-layer flexible codes and fixed EP codes under different storage constraint. $N=16,R=11$, and $\lambda=\kappa=\mu=1$ unit. Each server has a failure probability of $0.05$. The approximate computation load in \eqref{eq:optimal L_flex} and the expected computation load in \eqref{overall latency for 2 Layers} are both shown in the figure. For all cases, the optimal matrix partitioning and the recovery profile are found by exhaustive search.}\label{fig:2-layer comparison}
\end{figure}

Fig. \ref{fig:2-layer comparison} shows a comparison of our 2-layer flexible codes and the fixed EP codes. For the approximate computation load, only the computation load in the first layer is considered as in \eqref{optimal 2}.
The expected computation load is computed based on \eqref{overall latency for 2 Layers} with a truncated binomial distribution
\begin{align}\label{eq:binomial}
q_j = \frac{1}{\theta}\binom{N}{j}(1-\epsilon)^{N-j} \epsilon ^{j},  0 \le j \le N-R,
\end{align}
where $\epsilon = 0.05$ is the probability that each server is a straggler.
To limit the number of stragglers below $N-R$, we truncate the binomial distribution below $N-R$ and $\theta = \sum\limits_{i=0}^{N-R}{N \choose i}\epsilon^i (1-\epsilon)^{N-i}$ is the probability that there are at least $N-R$ available nodes.\footnote{In practice, $R$ is chosen such that the probability of having more than $N-R$ stragglers is negligible.} In Fig. \ref{fig:2-layer comparison}, we have $1 - \theta < 10^{-4}$. 
The minimum required storage constraint is $C=0.33$. When $C < 0.45$, our 2-layer construction reduces to the EP code. When the storage constraint $C \ge 0.45$, our 2-layer constructions have better performance. For example, when $C = 0.9$, the optimal EP code has $p_0 = 2,m_0 = 1,n_0 = 5$ and its expected computation is $L_{\text{EP}} = 0.1$. However, our 2-layer optimal flexible code has $p_1 = 1,m_1=3,n_1=5,p_2=6,m_2=1,n_2=1, R_1=15,$ and its expected computation load is $L_{\text{flex}} = 0.069$, i.e., we save more than $30\%$ in terms of computation load. In addition, the approximate computation load of the 2-layer flexible code in this case is $0.067$, which is very close to the expected computation load when the computation in both layers are considered.

\subsection{Optimization for the multi-round communication model}

Now let us consider the multi-round communication model where coded matrices are sent sequentially from the source to the server. In this case, it is only required that the maximum size of the  coded matrices does not exceed the storage size. As mentioned before Theorem \ref{thm: Construction 1}, the storage capacity just needs to exceed the size of the first pair of coded matrices. We first optimize the partitioning parameters for a fixed recovery profile and then optimize the number of layers and the recovery profile.

Let us consider the construction with $a$ layers and predetermined $R_j, j\in[a]$ such that $N\ge R_1 > R_2 > ... > R_a = R$.
Assuming $R_j,j\in[a],$ and the storage constraint $C$ are given, we first minimize the computation load in each layer: 
\begin{equation}\label{eq:multiple-layer optimazation}
	\begin{aligned}
		\min_{p_j,m_j,n_j} \quad &L_j\\
		\textrm{s.t.} \quad & R_j = p_jm_jn_j + p_j -1\\
		&\frac{\lambda\kappa}{p_1m_1} +  \frac{\kappa\mu}{p_1n_1} \leq C ,  \\
	\end{aligned}
\end{equation}
where $L_j$ is shown in \eqref{eq:general L_j}. Note that once $L_j, j \in [a],$ are minimized, by Theorem \ref{thm: Construction 1} the computation load $L_{\text{flex}}(\hat{R})$ for any number of non-stragglers $\hat{R}$ is also minimized. Hence the optimization in \eqref{eq:multiple-layer optimazation} is stronger than optimizing the expected computation load defined in \eqref{eq:expected_L}.

\begin{thm}\label{thm:multi_optimal_Lj}
	The optimal solution of \eqref{eq:multiple-layer optimazation} for the flexible construction under the multi-round communication model is
	\begin{align}
	    L_j^\ast = \begin{cases}
	    \frac{2C\lambda\kappa\mu}{C(R+1)+\sqrt{C^2(R+1)^2-16\lambda\kappa^2\mu}}, & j=1,\\
	    \frac{R_1(R_{j-1}-R_j)}{R_{j-1}R_j}L_1, & j \ge 2,
	    \end{cases}
	\end{align}
	with
	\begin{align}
		p_1^\ast = \frac{1}{2}(R+1) - \frac{1}{2}\sqrt{(R+1)^2-\frac{16\lambda\kappa^2\mu}{C^2}},
	\end{align}
	$m_1^\ast,n_1^\ast$ are given by $m_1^\ast n_1^\ast = \frac{R+1}{p_1^\ast} - 1$ and $\lambda\kappa n_1^\ast = \kappa\mu m_1^\ast$ and $p^\ast_j = 1, m^\ast_j n^\ast_j = R_j$ for $j\ge 2$.
\end{thm}

\begin{IEEEproof}
When $j = 1$, it is the same optimization problem in Theorem \ref{thm:optimazation 1}.

For $j \ge 2$, We prove by induction.

\textbf{Base case}: For $j=2$, since there is no constraint on storage size of Layer $2$, by Theorem \ref{thm: Construction 1} we have
\begin{align}
L_2=\frac{R_{1} - R_{2}}{p_2m_2n_2}L_1 =  \frac{R_{1} - R_{2}}{R_2-p_2+1}L_1,    
\end{align}
which is an increasing function of $p_2$. Thus, we have $p^\ast_2 = 1,m^\ast_2 n^\ast_2 = R_2$.

\textbf{Induction step}: Assume the minimum $L_J^\ast$ is achieved when $p^\ast_J = 1, m^\ast_J n^\ast_J = R_J$ for $J=2,3,...,j-1$. For $J=j$, we have
\begin{align}
	L_{j} = \frac{R_{j-1} - R_{j}}{p_jm_jn_j} \sum\limits_{J=1}^{j-1} L_{J}=\frac{R_{j-1} - R_{j}}{R_j-p_j+1}\sum\limits_{J=1}^{j-1} L_{J},
\end{align}
which is an increasing function of $p_j$ and $L_J, J \in [j-1]$, respectively. Hence, we should pick the minimum $p_j^\ast=1$ and the minimum $L_J^\ast, 1 \le J \le j-1$ to optimize $L_j$. Therefore, $p^\ast_J = 1, m^\ast_J n^\ast_J = R_J$ for all $2\le J \le j$. 
\end{IEEEproof}

Notice that in the case of $j\ge 2$, we have $R_j = m_jn_j$, there is at least one integer solution with $m_j = R_j,n_j = 1$, which simplifies the problem to be matrix-vector multiplication.

Next, we discuss how to set the number of layers and the recovery profile to minimize the computation load. First, we state a lemma to show that adding more layers does not increase the computation load. Then, a theorem is proposed to show how to set $R_1$.

\begin{lem} \label{lem:more_layer}
	Given $R, R_1$, and $p_j=1, j \ge 2$, adding another layer does not increase the computation load of each server.
\end{lem} 
\begin{IEEEproof}
	Let us add a layer between Layers $j-1$ and $j$.
	When $R_{j-1} = R_{j} + 1$, no layers can be added. When $R_{j-1} > R_{j} + 1$, consider adding one extra layer with $R_{\text{add}} = R_j+1$ between Layer $j-1$ and $j$ so that $R_{j} < R_{\text{add}} < R_{j-1}$. Denote the computation load of the new construction by $L_{\text{add}}$, which is a function of the number of non-stragglers, $\hat{R}$.
	
	When $\hat{R} \le R_{j}$ or $\hat{R} \ge R_{j-1}$, based on \eqref{eq:computation load in model 2}, the computation load of each server does not change, i.e., $L_{\text{add}} = L_{\text{flex}}$.
	
	When $R_{\text{add}} \leq \hat{R} < R_{j-1}$, by Corollary \ref{cor:optimal case}, the new computation load is
	\begin{align}
		L_{\text{add}} &=  \frac{R_1(R_{\text{add}} + R_{j-1}-\hat{R})}{R_{j-1}R_{\text{add}}}L_1 \nonumber \\
		& =\frac{R_1}{R_{j-1}}L_1 + \frac{R_1(R_{j-1}-\hat{R})}{R_{j-1}R_{\text{add}}}L_1,\nonumber \\
		& < \frac{R_1}{R_{j-1}}L_1 + \frac{R_1(R_{j-1}-\hat{R})}{R_{j-1}R_{j}}L_1 \nonumber \\
		& = L_{\text{flex}},
	\end{align}
	where the inequality results from the fact that $R_{\text{add}} > R_j$. Thus, by adding one layer with $R_{\text{add}} = R_{j}+1$, computation load does not increase. 
	Similarly, more layers can be added between Layer $j-1$ and $j$. Therefore, adding more layers between $R_1$ and $R$ does not increase the computation load of each server.
\end{IEEEproof}

Based on Lemma \ref{lem:more_layer}, given $R_1$ and $R$, the optimal scheme is to add one layer for each value between $R_1$ and $R$. Thus, the recovery profile should be chosen to be $(R_1,R_1-1,R_1-2,\dots,R)$. The only problem left is how to set $R_1$. 
According to Corollary \ref{cor:optimal case} and Theorem \ref{thm:multi_optimal_Lj},
\begin{align}\label{eq:ep}
	L_{\text{flex}} = 
	\begin{cases}
	L_1 = \frac{2\lambda\kappa\mu}{(R+1)+\sqrt{(R+1)^2-\frac{16\lambda\kappa^2\mu}{C^2}}}, & \mbox{if } \hat{R}>R_1,\\
	\frac{R_1}{N-j}L_1, & \mbox{if } \hat{R}= N-j, N-R_1 \le j \le N-R.	
	\end{cases}
\end{align}
Based on \eqref{eq:ep}, we see $\frac{\lambda\kappa\mu}{R_1+1} <  L_1 < \frac{2\lambda\kappa\mu}{R_1+1}$. Denote $L_1  = \eta \frac{\lambda\kappa\mu}{R_1+1}$, where 
\begin{align}
	\eta = \frac{2}{1+\sqrt{1-\frac{16\lambda\kappa^2\mu}{C^2(1+R_1)^2}}}.
\end{align}
Note that given $\lambda, \kappa, \mu, C$, the value of $\eta$ decreases as $R_1$ increases.
Then, for fixed $R_1$, the expectation of the computation load is 
\begin{align}
	E\left[L_{\text{flex}}^{(R_1)}\right] &= L_1 \sum_{j=0}^{N-R_1-1}q_j + \sum_{j=N-R_1}^{N-R}q_{j}\frac{R_1}{N-j}L_1 \\
	&= \lambda\kappa\mu \eta\left( \sum_{j=0}^{N-R_1-1}\frac{ q_j}{1+R_1} + \sum_{j=N-R_1}^{N-R} \frac{ q_{j} R_1}{(N-j)(1+R_1)} \right). \label{eq:expected_L_R1} 
\end{align}
Here, the superscript $R_1$ indicates that the computation load depends on $R_1$. The goal is to minimize $E\left[L_{\text{flex}}^{(R_1)}\right]$ over $R_1$ where $R \le R_1 \le N$. 

The theorem below states a sufficient condition for which we should set $R_1=N$ and use the maximum number of layers. In particular, the recovery profile should be $(N,N-1,N-2,\dots,R)$.

\begin{thm} \label{thm:R1}
	When $q_0 > \sum_{j=1}^{N-R} \frac{ q_{j}}{N-j}$, the optimal $R_1$ to minimize \eqref{eq:expected_L_R1} is achieved when $R_1^\ast=N$.
\end{thm}

\begin{IEEEproof}
Denote the term in the parentheses of \eqref{eq:expected_L_R1} as
\begin{align}
	h(R_1) =  \frac{1}{1+R_1}\sum_{j=0}^{N-R_1-1} q_j + \frac{R_1}{1+R_1}\sum_{j=N-R_1}^{N-R} \frac{ q_{j}}{N-j}.
\end{align}
When $R_1 = N$, 
\begin{align}
	h(N) =  \frac{N}{1+N}\sum_{j=0}^{N-R} \frac{ q_{j}}{N-j}.
\end{align}
When $R_1 = N-k, k \in[1, N-R]$,
\begin{align}
	h(N-k) = \frac{1}{N-k+1} \sum_{j=0}^{k-1}q_{j}  + \frac{N-k}{N-k+1}\sum_{j=k}^{N-R} \frac{ q_{j}}{N-j}.
\end{align}
Then, their difference is
\begin{align}
	&\qquad h(N-k) - h(N) \\
	&= \frac{1}{N-k+1} \sum_{j=0}^{k-1}q_{j}  + \frac{N-k}{N-k+1}\sum_{j=k}^{N-R} \frac{ q_{j}}{N-j}- \frac{N}{1+N}\sum_{j=0}^{N-R} \frac{ q_{j}}{N-j}\\
	& = \frac{1}{(N+1)(N-k+1)} \left( \sum_{j=0}^{k-1}\frac{(k-j)N-j}{N-j}q_{j} - k\sum_{j=k}^{N-R} \frac{ q_{j}}{N-j}\right). \label{eq:explict}
\end{align}
When $q_0 > \sum_{j=1}^{N-R} \frac{ q_{j}}{N-j}$, the first term in the parentheses of \eqref{eq:explict} is
\begin{align}
	\sum_{j=0}^{k-1}\frac{(k-j)N-j}{N-j}q_{j}  \ge kq_0 > k\sum_{j=1}^{N-R} \frac{ q_{j}}{N-j} \ge k\sum_{j=k}^{N-R} \frac{ q_{j}}{N-j}.
\end{align} 
Thus, $h(N-k)>h(N)$.
Since $\eta$ increases as $R_1$ decreases, by \eqref{eq:expected_L_R1} we conclude that $E\left[L_\text{flex}^{(N)}\right] < E\left[L_\text{flex}^{(N-k)}\right]$. Therefore, $R_1^\ast$ should be set as $N$.
\end{IEEEproof}

\begin{exa}
	Suppose $N=50, R=40$ and assume the number of stragglers follows a truncated binomial distribution similar to \eqref{eq:binomial}, i.e., $q_j=\theta{N \choose j}\epsilon^j (1-\epsilon)^{N-j}$, for the constant factor $\theta=\frac{1}{\sum_{i=0}^{N-R}{N \choose i}\epsilon^i (1-\epsilon)^{N-i}}$. 
	According to Theorem \ref{thm:R1}, $R_1$ can be set as $N$ as long as $\epsilon < 7.4\%$. 
\end{exa}

\section{Conclusion} \label{conclusion}
In this paper, we consider coded distributed matrix multiplication. A flexible construction for distributed matrix multiplication is proposed and the optimal parameters are discussed. The construction can also be generalized to batch processing of matrix multiplication and secure distributed computation. Flexible constructions are also found in other problems such as communication-efficient secret sharing \cite{Huang_efficientSS}, adaptive gradient codes \cite{malitsky2019adaptive}, coded elastic computing \cite{Yang_elastic} and flexible storage \cite{PatentHJ,Li_storage}. It is worthwhile to explore more applications of flexible constructions, such as distributed machine learning and secure multi-party computation.

	\clearpage
	\newpage
	\bibliographystyle{IEEEtran}
	\bibliography{IEEEabrv,sample}

\begin{thebibliography}{10}
\providecommand{\url}[1]{#1}
\csname url@samestyle\endcsname
\providecommand{\newblock}{\relax}
\providecommand{\bibinfo}[2]{#2}
\providecommand{\BIBentrySTDinterwordspacing}{\spaceskip=0pt\relax}
\providecommand{\BIBentryALTinterwordstretchfactor}{4}
\providecommand{\BIBentryALTinterwordspacing}{\spaceskip=\fontdimen2\font plus
\BIBentryALTinterwordstretchfactor\fontdimen3\font minus
  \fontdimen4\font\relax}
\providecommand{\BIBforeignlanguage}[2]{{%
\expandafter\ifx\csname l@#1\endcsname\relax
\typeout{** WARNING: IEEEtran.bst: No hyphenation pattern has been}%
\typeout{** loaded for the language `#1'. Using the pattern for}%
\typeout{** the default language instead.}%
\else
\language=\csname l@#1\endcsname
\fi
#2}}
\providecommand{\BIBdecl}{\relax}
\BIBdecl

\bibitem{Li_flexible}
W.~Li, Z.~Chen, Z.~Wang, S.~A. Jafar, and H.~Jafarkhani, ``Flexible
  constructions for distributed matrix multiplication,'' in \emph{Proceedings
  of International Symposium on Information Theory (ISIT)}, 2021.

\bibitem{Dean_tail}
J.~Dean and L.~A. Barroso, ``The tail at scale,'' \emph{Communications of the
  ACM}, vol.~56, pp. 74--80, 2013.

\bibitem{Anan_straggler}
G.~Ananthanarayanan, S.~Kandula, A.~Greenberg, I.~Stoica, Y.~Lu, B.~Saha, and
  E.~Harris, ``Reining in the outliers in map-reduce clusters using mantri,''
  in \emph{Proceedings of the 9th USENIX Conference on Operating Systems Design
  and Implementation}, ser. OSDI'10.\hskip 1em plus 0.5em minus 0.4em\relax
  USA: USENIX Association, 2010, p. 265–278.

\bibitem{Lee_Lam_Pedarsani}
K.~Lee, M.~Lam, R.~Pedarsani, D.~Papailiopoulos, and K.~Ramchandran, ``Speeding
  up distributed machine learning using codes,'' \emph{IEEE Transactions on
  Information Theory}, vol.~64, no.~3, pp. 1514--1529, 2017.

\bibitem{Yu_Maddah-Ali_Avestimehr_Polynomial}
Q.~Yu, M.~A. Maddah-Ali, and A.~S. Avestimehr, ``Polynomial codes: an optimal
  design for high-dimensional coded matrix multiplication,'' \emph{arXiv
  preprint arXiv:1705.10464}, 2017.

\bibitem{Dutta_Fahim_Haddadpour}
S.~Dutta, M.~Fahim, F.~Haddadpour, H.~Jeong, V.~Cadambe, and P.~Grover, ``On
  the optimal recovery threshold of coded matrix multiplication,'' \emph{IEEE
  Transactions on Information Theory}, vol.~66, no.~1, pp. 278--301, 2020.

\bibitem{GPolyDot}
S.~Dutta, Z.~Bai, H.~Jeong, T.~Low, and P.~Grover, ``A unified coded deep
  neural network training strategy based on generalized polydot codes for
  matrix multiplication,'' \emph{ArXiv:1811.10751}, Nov. 2018.

\bibitem{Yu_Maddah-Ali_Avestimehr}
Q.~Yu, M.~A. Maddah-Ali, and A.~S. Avestimehr, ``Straggler mitigation in
  distributed matrix multiplication: Fundamental limits and optimal coding,''
  \emph{IEEE Transactions on Information Theory}, vol.~66, no.~3, pp.
  1920--1933, 2020.

\bibitem{Yu_Lagrange}
Q.~Yu, S.~Li, N.~Raviv, S.~M.~M. Kalan, M.~Soltanolkotabi, and S.~Avestimehr,
  ``Lagrange coded computing: Optimal design for resiliency, security and
  privacy,'' \emph{ArXiv:1806.00939}, 2018.

\bibitem{Reisizadeh_Prakash_Pedarsani}
A.~Reisizadeh, S.~Prakash, R.~Pedarsani, and A.~S. Avestimehr, ``Coded
  computation over heterogeneous clusters,'' \emph{IEEE Transactions on
  Information Theory}, vol.~65, no.~7, pp. 4227--4242, 2019.

\bibitem{Lee_Suh_Ramchandran}
K.~Lee, C.~Suh, and K.~Ramchandran, ``High-dimensional coded matrix
  multiplication,'' in \emph{2017 IEEE International Symposium on Information
  Theory (ISIT)}.\hskip 1em plus 0.5em minus 0.4em\relax IEEE, 2017, pp.
  2418--2422.

\bibitem{Dutta_Cadambe_Short}
S.~Dutta, V.~Cadambe, and P.~Grover, ``Short-dot: Computing large linear
  transforms distributedly using coded short dot products,'' in \emph{Advances
  In Neural Information Processing Systems}, 2016, pp. 2100--2108.

\bibitem{Dutta_Cadambe_Codedconv}
------, ``Coded convolution for parallel and distributed computing within a
  deadline,'' \emph{arXiv preprint arXiv:1705.03875}, 2017.

\bibitem{Yu_Maddah-Ali_CodedDFT}
Q.~Yu, M.~A. Maddah-Ali, and A.~S. Avestimehr, ``Coded fourier transform,''
  \emph{arXiv preprint arXiv:1710.06471}, 2017.

\bibitem{Jahani-Nezhad_Maddah-Ali}
T.~Jahani-Nezhad and M.~A. Maddah-Ali, ``Codedsketch: A coding scheme for
  distributed computation of approximated matrix multiplications,'' \emph{arXiv
  preprint arXiv:1812.10460}, 2018.

\bibitem{Baharav_Lee_Ocal}
T.~Baharav, K.~Lee, O.~Ocal, and K.~Ramchandran, ``Straggler-proofing
  massive-scale distributed matrix multiplication with d-dimensional product
  codes,'' in \emph{2018 IEEE International Symposium on Information Theory
  (ISIT)}.\hskip 1em plus 0.5em minus 0.4em\relax IEEE, 2018, pp. 1993--1997.

\bibitem{Suh_Lee_Msparse}
G.~Suh, K.~Lee, and C.~Suh, ``Matrix sparsification for coded matrix
  multiplication,'' in \emph{2017 55th Annual Allerton Conference on
  Communication, Control, and Computing (Allerton)}.\hskip 1em plus 0.5em minus
  0.4em\relax IEEE, 2017, pp. 1271--1278.

\bibitem{Wang_Liu_CLT}
S.~Wang, J.~Liu, N.~Shroff, and P.~Yang, ``Fundamental limits of coded linear
  transform,'' \emph{arXiv preprint arXiv:1804.09791}, 2018.

\bibitem{Mallick_Chaudhari_Joshi}
A.~Mallick, M.~Chaudhari, U.~Sheth, G.~Palanikumar, and G.~Joshi, ``Rateless
  codes for near-perfect load balancing in distributed matrix-vector
  multiplication,'' \emph{Proc. ACM Meas. Anal. Comput. Syst.}, vol.~3, no.~3,
  2019.

\bibitem{Wang_Liu_Sparse}
S.~Wang, J.~Liu, and N.~Shroff, ``Coded sparse matrix multiplication,''
  \emph{arXiv preprint arXiv:1802.03430}, 2018.

\bibitem{Severinson_iAmat_Rosnes}
A.~Severinson, A.~G. i~Amat, and E.~Rosnes, ``Block-diagonal and lt codes for
  distributed computing with straggling servers,'' \emph{IEEE Transactions on
  Communications}, vol.~67, no.~3, pp. 1739--1753, 2018.

\bibitem{Haddadpour_Cadambe_Finite}
F.~Haddadpour and V.~R. Cadambe, ``Codes for distributed finite alphabet
  matrix-vector multiplication,'' in \emph{2018 IEEE International Symposium on
  Information Theory (ISIT)}.\hskip 1em plus 0.5em minus 0.4em\relax IEEE,
  2018, pp. 1625--1629.

\bibitem{Sheth_Dutta_Chaudhari}
U.~Sheth, S.~Dutta, M.~Chaudhari, H.~Jeong, Y.~Yang, J.~Kohonen, T.~Roos, and
  P.~Grover, ``An application of storage-optimal matdot codes for coded matrix
  multiplication: Fast k-nearest neighbors estimation,'' in \emph{2018 IEEE
  International Conference on Big Data (Big Data)}.\hskip 1em plus 0.5em minus
  0.4em\relax IEEE, 2018, pp. 1113--1120.

\bibitem{Jeong_Ye_Grover}
H.~Jeong, F.~Ye, and P.~Grover, ``Locally recoverable coded matrix
  multiplication,'' in \emph{2018 56th Annual Allerton Conference on
  Communication, Control, and Computing (Allerton)}.\hskip 1em plus 0.5em minus
  0.4em\relax IEEE, 2018, pp. 715--722.

\bibitem{Kim_Sohn_Moon_Group}
M.~Kim, J.-y. Sohn, and J.~Moon, ``Coded matrix multiplication on a group-based
  model,'' \emph{arXiv preprint arXiv:1901.05162}, 2019.

\bibitem{Park_Lee_Sohn}
H.~Park, K.~Lee, J.-y. Sohn, C.~Suh, and J.~Moon, ``Hierarchical coding for
  distributed computing,'' \emph{arXiv preprint arXiv:1801.04686}, 2018.

\bibitem{Li_Maddah-Ali_Fog}
S.~Li, M.~A. Maddah-Ali, and A.~S. Avestimehr, ``Coding for distributed fog
  computing,'' \emph{IEEE Communications Magazine}, vol.~55, no.~4, pp. 34--40,
  2017.

\bibitem{Chang_Tandon}
W.~Chang and R.~Tandon, ``On the capacity of secure distributed matrix
  multiplication,'' \emph{IEEE Global Communications Conference (GLOBECOM)},
  pp. 1--6, 2018.

\bibitem{Kakar_Ebadifar_Sezgin_CSA}
J.~Kakar, S.~Ebadifar, and A.~Sezgin, ``On the capacity and
  straggler-robustness of distributed secure matrix multiplication,''
  \emph{IEEE Access}, vol.~7, pp. 45\,783--45\,799, 2019.

\bibitem{Oliveira_Rouayheb_Karpuk}
R.~G. D’Oliveira, S.~E. Rouayheb, and D.~Karpuk, ``Gasp codes for secure
  distributed matrix multiplication,'' \emph{IEEE Transactions on Information
  Theory}, 2020, early access, DOI: 10.1109/TIT.2020.2975021.

\bibitem{Kim_Lee}
M.~Kim and J.~Lee, ``Private secure coded computation,'' \emph{IEEE
  Communications Letters}, vol.~23, no.~11, pp. 1918--1921, 2019.

\bibitem{Aliasgari_Simeone_Kliewer}
M.~Aliasgari, O.~Simeone, and J.~Kliewer, ``Distributed and private coded
  matrix computation with flexible communication load,'' \emph{arXiv preprint
  arXiv:1901.07705}, 2019.

\bibitem{Jia_Jafar_CDBC}
Z.~Jia and S.~Jafar, ``Cross-subspace alignment codes for coded distributed
  batch computation,'' \emph{ArXiv:1909.13873}, 2019.

\bibitem{Chen_Jia_Wang_Jafar_GCSANA}
Z.~Chen, Z.~Jia, Z.~Wang, and S.~A. Jafar, ``Gcsa codes with noise alignment
  for secure coded multi-party batch matrix multiplication,'' \emph{IEEE
  Journal on Selected Areas in Information Theory, Early Access}, 2021.

\bibitem{Nuwan_Stark_ISIT}
N.~Ferdinand and S.~C. Draper, ``Hierarchical coded computations,'' \emph{IEEE
  International Symposium on Information Theory}, 2018.

\bibitem{Amiri_cdc}
M.~M. Amiri and D.~G\"{u}nd\"{u}z, ``Computation scheduling for distributed
  machine learning with straggling workers,'' \emph{IEEE Transactions on Signal
  Processing}, vol.~67, no.~24, pp. 6270--6284, 2019.

\bibitem{Bitar_Parag_Rouayheb_staircasecodes}
R.~Bitar, P.~Parag, and S.~E. Rouayheb, ``Minimizing latency for secure coded
  computing using secret sharing via staircase codes,'' \emph{IEEE Transactions
  on Communications}, vol.~68, no.~8, pp. 4609--4619, 2020.

\bibitem{Bitar_Xing_adaptive}
R.~Bitar, Y.~Xing, Y.~Keshtkarjahromi, V.~Dasari, S.~E. Rouayheb, and
  H.~Seferoglu, ``Private and rateless adaptive coded matrix-vector
  multiplication,'' \emph{arXiv preprint arXiv:1909.12611}, 2019.

\bibitem{Ramamoorthy_universal}
A.~Ramamoorthy, L.~Tang, and P.~O. Vontobel, ``Universally decodable matrices
  for distributed matrix-vector multiplication,'' \emph{arXiv:1901.10674},
  2019.

\bibitem{Das_c3les}
A.~B. Das, L.~Tang, and A.~Ramamoorthy, ``C3les: Codes for coded computation
  that leverage stragglers,'' in \emph{2018 IEEE Information Theory Workshop
  (ITW)}, 2018, pp. 1--5.

\bibitem{Bitar_Xhemrishi_adaptive}
R.~Bitar, M.~Xhemrishi, and A.~Wachter-Zeh, ``Adaptive private distributed
  matrix multiplication,'' \emph{arXiv preprint arXiv:2101.05681}, 2021.

\bibitem{Kiani_Nuwan_Stark_ISIT}
S.~Kiani, N.~Ferdinand, and S.~C. Draper, ``Exploitation of stragglers in coded
  computation,'' \emph{IEEE International Symposium on Information Theory},
  2018.

\bibitem{Hasircioglu_bivariate}
B.~Has{\i}rc{\i}oglu, J.~G\'{o}mez-Vilardeb\'{o}, and D.~G\"{u}nd\"{u}z,
  ``Bivariate polynomial coding for exploiting stragglers in heterogeneous
  coded computing systems,'' \emph{ArXiv:2001.07227}, 2020.

\bibitem{Hasircioglu_Hermitian}
B.~Has{\i}rc{\i}o\u{g}lu, J.~G\'{o}mez-Vilardeb\'{o}, and D.~G\"{u}nd\"{u}z,
  ``Bivariate hermitian polynomial coding for efficient distributed matrix
  multiplicationn,'' \emph{2020 IEEE Global Communications Conference}, pp.
  1--6, 2020.

\bibitem{Fan_partial}
X.~Fan, P.~Soto, X.~Zhong, D.~Xi, Y.~Wang, and J.~Li, ``Leveraging stragglers
  in coded computing with heterogeneous servers,'' in \emph{2020 IEEE/ACM 28th
  International Symposium on Quality of Service (IWQoS)}, 2020, pp. 1--10.

\bibitem{Das_sparse}
A.~B. Das and A.~Ramamoorthy, ``Coded sparse matrix computation schemes that
  leverage partial stragglers,'' \emph{arXiv:2012.06065}, 2020.

\bibitem{Shahrzad_Hierarchical}
S.~Kianidehkordi, N.~Ferdinand, and S.~C. Draper, ``Hierarchical coded matrix
  multiplication,'' \emph{IEEE Transactions on Information Theory}, vol.~67,
  no.~2, pp. 726--754, 2021.

\bibitem{gashkov2013complexity}
S.~B. Gashkov and I.~S. Sergeev, ``Complexity of computation in finite
  fields,'' \emph{Journal of Mathematical Sciences}, vol. 191, no.~5, pp.
  661--685, 2013.

\bibitem{sathiamoorthy2013xoring}
M.~Sathiamoorthy, M.~Asteris, D.~Papailiopoulos, A.~G. Dimakis, R.~Vadali,
  S.~Chen, and D.~Borthakur, ``Xoring elephants: Novel erasure codes for big
  data,'' in \emph{39th Int. Conf. Very Large Data Bases}, vol.~6, no.~5, 2013,
  pp. 325--336.

\bibitem{Huang_efficientSS}
W.~Huang, M.~Langberg, J.~Kliewer, and J.~Bruck, ``Communication efficient
  secret sharing,'' \emph{IEEE Transactions on Information Theory}, vol.~62,
  no.~12, pp. 7195--7206, 2016.

\bibitem{malitsky2019adaptive}
Y.~Malitsky and K.~Mishchenko, ``Adaptive gradient descent without descent,''
  \emph{arXiv preprint arXiv:1910.09529}, 2019.

\bibitem{Yang_elastic}
Y.~Yang, M.~Interlandi, P.~Grover, S.~Kar, S.~Amizadeh, and M.~Weimer, ``Coded
  elastic computing,'' \emph{arXiv preprint arXiv:1812.06411}, 2018.

\bibitem{PatentHJ}
H.~Jafarkhani and M.~Hajiaghayi, ``Cost-efficient repair for storage systems
  using progressive engagement,'' US Patent 10,187,088., Jan. 2019.

\bibitem{Li_storage}
W.~Li, Z.~Wang, T.~Lu, and H.~Jafarkhani, ``Storage codes with flexible number
  of nodes,'' \emph{arXiv preprint arXiv:2106.11336}, 2021.

\end{thebibliography}
\end{document}